\documentclass[
  aps,
  pra,
  superscriptaddress,
  reprint
]{revtex4-2}

\usepackage{array}[=2016-10-06]
\usepackage[T1]{fontenc}
\usepackage[american]{babel}
\usepackage{newtxtext}

\usepackage{amsmath}
\usepackage{amssymb}
\usepackage{mathtools}
\usepackage{newtxmath}
\usepackage{xfrac}
\usepackage{subdepth}
\usepackage{physics}
\usepackage{mleftright}
\mleftright
\usepackage{bm}

\usepackage{dcolumn}
\usepackage{scalerel}
\usepackage{comment}
\usepackage{graphicx}
\usepackage[HTML]{xcolor}

\usepackage{circledsteps}
\usepackage{booktabs}
\usepackage{tabularx}
\usepackage{etoolbox}
\usepackage[normalem]{ulem}

\usepackage{orcidlink}
\usepackage{hyperref}
\usepackage[all]{hypcap}
\usepackage{adjustbox}
\usepackage{microtype}
\newcolumntype{Y}{>{\raggedright\arraybackslash}X}

\DeclareMathOperator{\sgn}{sgn}

\definecolor{linkblue}{HTML}{2E3092}

\hypersetup{
  colorlinks = true,
  linkcolor  = linkblue,
  citecolor  = linkblue,
  urlcolor   = linkblue,
  filecolor  = linkblue
}

\allowdisplaybreaks

\newcommand{\PRLsep}{\begin{center}\resizebox{0.75\linewidth}{1.5pt}{$\bullet$}\end{center}}

\begin{document}
\let\origaddtocontents\addtocontents
\renewcommand{\addtocontents}[2]{%
  \ifstrequal{#1}{toc}{}{\origaddtocontents{#1}{#2}}%
}

\makeatletter
\patchcmd{\dch@set@one}{H_2}{H_b}
  {}{\errmessage{Subdepth patch 1 failed}}
\patchcmd{\dch@set@one}{H_2}{H_b}
  {}{\errmessage{Subdepth patch 2 failed}}
\patchcmd{\dch@set@one}{+\vrule}{\dagger\vrule}
  {}{\errmessage{Subdepth patch 3 failed}}
\makeatother

\title{Multicomponent anyons in one-dimensional optical lattices}

\author{Sagarika Basak\,\orcidlink{0000-0003-2069-644X}}
\email[\!Contact author:~]{basak.sagarika@rice.edu}
\altaffiliation{\href{mailto:basak.sagarika@ou.edu}{basak.sagarika@ou.edu}}
\affiliation{Department of Physics and Astronomy, and the Smalley-Curl Institute, \href{https://ror.org/008zs3103}{Rice University}, Houston, Texas 77251-1892, USA}
\affiliation{%
\adjustbox{max width=\textwidth}{%
Homer L. Dodge Department of Physics and Astronomy,
\href{https://ror.org/02aqsxs83}{The University of Oklahoma},
Norman, Oklahoma 73019, USA%
}}
\affiliation{Center for Quantum Research and Technology, \href{https://ror.org/02aqsxs83}{The University of Oklahoma}, Norman, Oklahoma 73019, USA}

\author{Xi-Wen Guan\,\orcidlink{0000-0001-6293-8529}}
\email[\!Contact author:~]{xiwen.guan@anu.edu.au}
\affiliation{%
\adjustbox{max width=\textwidth}{%
Innovation Academy for Precision Measurement Science and Technology,
\href{https://ror.org/034t30j35}{Chinese Academy of Sciences},
Wuhan 430071, China%
}}
\affiliation{%
\adjustbox{max width=0.99\textwidth}{%
Department of Fundamental and Theoretical Physics, Research School of Physics,
\href{https://ror.org/019wvm592}{Australian National University},
Canberra, ACT 0200, Australia%
}}

\author{Han Pu\,\orcidlink{0000-0002-0018-3076}}
\email[\!Contact author:~]{hpu@rice.edu}
\affiliation{Department of Physics and Astronomy, and the Smalley-Curl Institute, \href{https://ror.org/008zs3103}{Rice University}, Houston, Texas 77251-1892, USA}


\begin{abstract}
We investigate the ground-state and dynamical properties of multicomponent interacting anyons confined in a one-dimensional (1D) optical lattice. Adopting the Anyon--Hubbard model, we explore spin and charge correlations as functions of interaction strength and the anyonic statistical parameter. Our results demonstrate that multicomponent effects combined with fractional exchange statistics substantially reshape both charge and spin correlations. For fractional exchange statistics, the canonical symmetric shell structure with prominent singularities in the spinor fermionic momentum distribution undergoes notable structural reconstruction, exhibiting emergent asymmetry, new spectral peaks, and broadened singular features. Unlike the Tonks--Girardeau bosonic gas, pseudobosons, representing a limiting case of anyons, cease to display a dominant zero-momentum peak. Instead, a quasi-fermionic shell structure emerges with finite-momentum peaks, whose locations are modulated by lattice site occupancy and interaction strength. The structure factor of spin correlations uncovers antiferromagnetic ordering in spinor anyon systems, with correlation magnitudes tunable by anyonic exchange statistics. Furthermore, quench dynamics analysis reveals the statistical phase parameter as an effective tuning knob. It enables the switching of dipole oscillations between underdamped and overdamped relaxation regimes and governs overall cloud expansion dynamics. Our findings pave the way for exploring statistics-driven ground-state and dynamical phase transitions in spinor anyon systems.
\end{abstract}




\maketitle

\section{\label{sec:level1}Introduction}

A collection of identical particles can be classified based on their quantum statistics or, equivalently, the behavior of the system’s wave function under the exchange of two particles. Traditionally, there are two mutually exclusive groups: bosons, when the wave function is symmetric under exchange, and fermions, when the wave function is antisymmetric. In two dimensions, a third category of particles, \textit{anyons}, admits fractional exchange statistics, with the statistical phase able to take any value between $0$ and $\pi$ \cite{leinaas_theory_1977,wilczek_quantum_1982,wilczek_fractional_1990}. Two-dimensional systems of anyons have garnered a lot of attention \cite{wilczek_quantum_1982,haldane_fractional_1991,wilczek_fractional_1990,laughlin_anomalous_1983,halperin_statistics_1984,camino_realization_2005}, playing a crucial role in condensed matter physics \cite{wu_bosonization_1995,ha_exact_1994,murthy_thermodynamics_1994,kim_signatures_2005} and explaining the properties of fractional quantum Hall states \cite{arovas_fractional_1984,halperin_statistics_1984,bartolomei_fractional_2020,nakamura_direct_2020} or spin liquids \cite{yao_exact_2007}. Recently, anyons have also found utility in quantum computation \cite{lahtinen_short_2017}, enabling the buildup of massive collective memory and forming the building blocks for the exploration of topological quantum computation \cite{kitaev_fault-tolerant_2003,lahtinen_short_2017,bravyi_universal_2006,clarke_exotic_2013, google_quantum_ai_and_collaborators_non-abelian_2023}.

A generalization of the Pauli exclusion principle introduced a dimension-independent approach to anyonic statistics \cite{haldane_fractional_1991}, allowing the study of anyons in one dimension (1D). 1D systems, being governed by collective excitations, demonstrate enhanced quantum effects that lead to distinctive behaviors such as ``fermionization'' of bosons \cite{girardeau_relationship_1960,Tonks1936,lieb_exact_1963,Lieb1963b} and spin--charge separation in interacting fermions \cite{hilker_revealing_2017,auslaender_tunneling_2002,auslaender_spin-charge_2005,jompol_probing_2009,segovia_observation_1999, kim_observation_1996,kim_distinct_2006,senaratne_SCS_2022}. The study of 1D anyons was initiated in Refs.~\cite{kundu_exact_1999,batchelor_generalized_2006} and has gained immense interest \cite{girardeau_anyon-fermion_2006,calabrese_correlation_2007,hao_ground-state_2009,hao_dynamical_2012,wang_quantum_2014,averin_coulomb_2007,feiguin_interacting_2007,patu_one-dimensional_2008,patu_large-distance_2009,patu_one-dimensional_2009,santos_quantum_2012,bellazzini_junctions_2009,santachiara_one-particle_2008,greiter_statistical_2009,fidkowski_c_2008,trebst_collective_2008}, with great progress in theoretical tools \cite{amico_one-dimensional_1998,kundu_exact_1999,santachiara_one-particle_2008,patu_one-dimensional_2008,zhu_topological_1996} leading to the discovery of rich ground states \cite{arcila-forero_critical_2016,lange_anyonic_2017,tang_ground-state_2015,hao_ground-state_2009,patu_correlation_2015,zinner_strongly_2015,hao_ground-state_2008} and dynamics \cite{hao_dynamical_2012, wang_quantum_2014,del_campo_fermionization_2008,lau_quantum_2022,piroli_exact_2017,Tesfaye2026}. Exploration of exact solutions and mappings \cite{kundu_exact_1999,girardeau_anyon-fermion_2006,batchelor_bethe_2007,batchelor_inverse-scattering_2008,hidalgosacoto_twoanyons_2025,WangChenCui2025}, correlation functions \cite{wang_quantum_2014, hao_ground-state_2008,hao_ground-state_2009,patu_correlation_2007,patu_correlation_2019,tang_ground-state_2015,calabrese_correlation_2007,hidalgosacoto_universal_2025}, ground-state phases \cite{bonkhoff_bosonic_2021,lange_anyonic_2017,greschner_anyon_2015,arcila-forero_critical_2016}, phase transitions \cite{santos_quantum_2012,keilmann_statistically_2011,arcila-forero_critical_2016,zhang_ground-state_2017,Bonkhoff2025PhaseTransitions} and fermionization \cite{hao_dynamical_2012} are some of the key developments in the analysis of the anyonic ground state. Investigations of anyonic dynamics have led to the examination of dynamical fermionization and bosonization \cite{del_campo_fermionization_2008,Patu2025Bosonization}, relaxation dynamics \cite{hao_quench_2023,wright_nonequilibrium_2014,hao_dynamical_2012,wang_quantum_2014}, and nonequilibrium dynamics at finite temperature \cite{patu_nonequilibrium_2020}. Cold atomic setups, with the progress in experimental trapping and control \cite{paredes_tonksgirardeau_2004,kinoshita_observation_2004,haller_realization_2009,moritz_confinement_2005,liao_spin-imbalance_2010} and with the potential for tuning the quantum statistics \cite{leinaas_theory_1977,wilczek_quantum_1982,wilczek_fractional_1990}, provide a perfect platform for the realization and examination of 1D anyons. This has led to theoretical proposals such as lattice-shaking-induced resonant tunneling \cite{strater_floquet_2016} and multicolor lattice depth modulation \cite{cardarelli_engineering_2016,greschner_probing_2018}, as well as experimental realization \cite{kwan_realization_2024,bakkali_anyons_2026} of 1D anyons using cold atoms. Emergent many-body anyonic correlations were generated and probed in a strongly interacting 1D quantum gas via spin--charge separation and a mobile impurity \cite{dhar_expcorr_2025,Wang2025Swap}.

Consideration of multiple components diversifies the internal structure and provides additional ``spin'' degrees of freedom, resulting in complex system behavior such as spin-incoherent Luttinger liquids \cite{dutta_non-standard_2015}, ``flavor-selective'' Mott insulators \cite{vojta_orbital-selective_2010,tusi_flavour-selective_2022}, and a rich spin phase diagram under competing system parameters \cite{murmann_antiferromagnetic_2015,mazurenko_cold-atom_2017,messer_exploring_2015,basak_strongly_2021}. Multicomponent anyons in the continuum setup are proposed and investigated using a generalization of the anyonic Lieb--Liniger model \cite{santos_quantum_2012,patu_correlation_2019,patu_nonequilibrium_2023} and in an optical lattice examining the density expansion dynamics with deformed exchange statistics \cite{greschner_probing_2018}, while recent work has predicted asymmetric and chiral transport in two-component anyon–Hubbard systems with synthetic gauge flux \cite{ChenHuangZhangZhang2026}. By providing a natural setting in which internal degrees of freedom, interactions, and fractional exchange statistics can compete, multicomponent anyons hold considerable potential for rich physics. This raises a central question: how does fractional exchange statistics reorganize correlations and control transport when internal degrees of freedom are present?

Here, we study the ground-state and transport properties of multicomponent anyons in a 1D optical lattice. We ask how fractional exchange statistics modifies charge and spin correlations as the interaction strength, filling, and number of internal components are varied, and how these equilibrium signatures carry over to nonequilibrium transport. We characterize the ground state through the momentum distribution and spin structure factor and examine two complementary dynamical protocols: a sudden displacement of the harmonic trap and a release of a localized impurity. Together, these protocols probe how the statistical phase controls collective motion, impurity propagation, and momentum-space dynamics.
    
The remainder of this paper is organized as follows. Section~\ref{sec:thesystem} describes the system of multicomponent (spinor) anyons trapped in a 1D optical lattice, its generalized exclusion statistics, and its mapping to spinor fermions with occupation-dependent tunneling. Section~\ref{sec:results} describes the ground-state properties of spinor anyons. Section~\ref{sec:timeev} examines the transport properties: dipole oscillations illustrate the predominantly charge-sector response, and the localized impurity release probes the statistics-dependent impurity transport. Finally, Sec.~\ref{sec:summary} provides a summary and outlook. Details of the technical derivations and additional background information are relegated to Appendices~\ref{sec:appendA}--\ref{sec:appendG}.

\section{\label{sec:thesystem}System}

\begin{figure}
\centering
\includegraphics{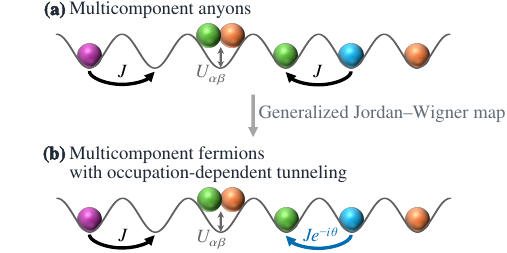}
\caption{\label{fig:schematic}\textbf{Fermionic representation of multicomponent anyons.}
(a) Nearest-neighbor tunneling with amplitude $J$ at on-site intercomponent interaction $U_{\alpha\beta}$. Colors denote the $N$ internal components.
(b) The generalized Jordan--Wigner transformation maps the anyons onto multicomponent fermions with direction- and occupation-dependent Peierls-type tunneling $J\exp(\mp i\theta n_i)$.}	
\end{figure}

We consider a system of $N$-component anyons trapped in a 1D optical lattice in the tight-binding limit at low temperature \hyperref[fig:schematic]{[Fig.~\ref{fig:schematic}(a)]}. This describes an $N$-component Anyon--Hubbard model, governed by the Hamiltonian \cite{osterloh_fermionic_2000,greschner_probing_2018}
\begin{align}
\label{eqn:anyonH}
H_{\mathrm{A}} = -J\sum\limits_{i,\alpha}(a_{i,\alpha}^{\dagger}a_{i+1,\alpha} + \text{H.c.})
+\sum\limits_{i,\alpha<\beta} U_{\alpha\beta}n_{i,\alpha}n_{i,\beta}\,,
\end{align}
where $J$ is the tunneling coefficient, $a_{i,\alpha}$ the anyonic annihilation operator for spin component $\alpha$ at lattice site $i$, and $U_{\alpha\beta}$ the on-site $\alpha\leftrightarrow\beta$ intercomponent interaction, $n_{i,\alpha}=a^\dagger_{i,\alpha}a_{i,\alpha}$ and $n_i=\sum_\alpha n_{i,\alpha}$. Spinor anyons follow the generalized exclusion statistics \cite{patu_correlation_2019}
\begin{align}
\label{eqn:anyonC}
\begin{split}
&a_{j,\alpha}a_{k,\beta}^{\dagger} +e^{-i\theta \sgn(j-k)} a_{k,\beta}^{\dagger}a_{j,\alpha} = \delta_{j,k}\delta_{\alpha,\beta}\,,
\\& a_{j,\alpha}a_{k,\beta} + e^{+i\theta \sgn(j-k)} a_{k,\beta}a_{j,\alpha}=0 \,.
\end{split}
\end{align}
$\theta$ is a statistical phase that anyons acquire upon exchange. For $\theta=0$, the particles act as fermions, whereas for $\theta=\pi$ they act as pseudobosons. The particles anticommute on the same site and obey generalized commutation relations on different sites. $\theta$ is considered spin independent and the same for all components. It is generalizable to component-dependent exchange phases, but that will be a future consideration.

\paragraph*{Fermionic mapping.}Multicomponent anyons are mapped to multicomponent fermions via a generalized Jordan--Wigner transformation \cite{keilmann_statistically_2011}:
\begin{align}
\label{eqn:JWMap}
a_{i,\alpha} = \exp\left(-i\theta \sum\nolimits_{l<i}n_l\right)c_{i,\alpha}\,,
\end{align}
where $c_{i,\alpha}$ annihilates a local fermion of component $\alpha$ at site $i$. The Jordan--Wigner string makes the representation of the anyonic operator nonlocal and ensures the generalized commutation relations in Eq.~\eqref{eqn:anyonC}; see Appendix~\ref{sec:appendA}.

The mapped fermions \hyperref[fig:schematic]{[Fig.~\ref{fig:schematic}(b)]} are governed by the generalized multicomponent Fermi--Hubbard Hamiltonian obtained by applying Eq.~\eqref{eqn:JWMap} to Eq.~\eqref{eqn:anyonH}:
\begin{align}
\label{eqn:fermiH}
H_{\mathrm{F}} = 
&-J\sum\limits_{i,\alpha} \left(c_{i,\alpha}^{\dagger}e^{-i\theta n_{i}}c_{i+1,\alpha} + c_{i+1,\alpha}^{\dagger}e^{+i\theta n_{i}} c_{i,\alpha}\right) 
\nonumber\\&+\smashoperator{\sum\limits_{i,\alpha<\beta}} U_{\alpha\beta}n_{i,\alpha}n_{i,\beta}\,.
\end{align}
This Hamiltonian features occupation-dependent tunneling $\left[J\exp(\pm i\theta n_{i})\right]$ with a Peierls-type tunneling phase that breaks the reflection parity appearing in the tunneling coefficient. The generalized Jordan--Wigner transformation therefore maps a system of anyons hopping between the nearest-neighbor sites to a system of fermions with occupation-dependent tunneling between the nearest-neighbor sites. Density operators remain unaffected in this mapping; the interaction term in the Hamiltonian is thus the same for fermions and anyons. Although the mapping is nonlocal, the Hamiltonian here has only local operators. Based on the recent proposals and realization of anyons using cold atoms \cite{cardarelli_engineering_2016, kwan_realization_2024}, multicomponent anyons can potentially be realized using fermions in a tilted lattice with tunneling induced by multicolor modulation of the lattice depth. Here, we have considered the anyons to anticommute on the same site, realizing a mapped system of fermions with occupation-dependent tunneling. If the anyons commute on the same site, they can instead be mapped to a system of bosons, with occupation-dependent tunneling and on-site inter- and intra-component interaction~\cite{basak_anyon2}. Hereon, we consider component-independent interaction $U_{\alpha\beta} = U$.

\section{\label{sec:results}Ground-state properties}
Exploring the ground state of \textit{single}-component (spinless) anyons has provided a wealth of new physics, such as statistically induced phase transitions \cite{keilmann_statistically_2011, santos_quantum_2012}, a novel superfluid phase \cite{greschner_anyon_2015}, and an asymmetric momentum distribution \cite{hao_ground-state_2008}. Introducing \textit{multiple components} adds internal spin degrees of freedom, allowing fractional exchange statistics to potentially affect both charge and spin correlations. Here, we analyze the ground state of multicomponent anyons using the Anyon--Hubbard Hamiltonian to uncover physics emerging from the spinor nature of these anyons. The impact of fractional exchange statistics and multiple components on charge and spin correlations is investigated. At a fixed total number of particles, the ground state is allowed to select both the spin configuration and the population of each component. Spin balance thus emerges as a ground-state property rather than an imposed constraint.

\subsection{Observables}
We will focus on two important experimentally measurable properties: 1.~Momentum distribution and 2.~Spin structure factor, obtained via density-matrix renormalization group (DMRG) for open boundary conditions (developed with the aid of Refs.~\cite{garrison_simple-dmrgsimple-dmrg_2017, schollwock_density-matrix_2011} and TeNPy implementation \cite{tenpy2024}).

\subsubsection{Momentum distribution} 
The intrinsic anyonic momentum distribution is defined as the expectation of the anyonic occupation operator in momentum space and expressed, by setting the lattice spacing to $1$, as \cite{manmana_su_2011}
\begin{align}
    n_k =  \dfrac{1}{N}\sum\limits_{\sigma} \expval{a_{k,\sigma}^{\dagger}a_{k,\sigma}} = \dfrac{1}{L}\sum\limits_{m,n=1}^{L} C(m,n) e^{ik(m-n)}\,,
\end{align}
where $N$ and $L$ denote the number of components and lattice sites, respectively. $n_k$ is the Fourier transform of the one-body density correlation $C(m,n)$, defined for spinor anyons as
\begin{align}
    C(m,n) &= \frac{1}{N}\sum_{\sigma}C_{\sigma}(m,n) \notag\\[2.5ex]&=  \frac{1}{N}\sum_{\sigma}\expval{a_{m,\sigma}^{\dagger}a_{n,\sigma}}\notag\\[2.5ex]
     &=
\begin{cases}
\begin{aligned}
    \frac{1}{N}&\sum\limits_{\sigma}\expval{c_{m,\sigma}^{\dagger}\exp\left(-i\theta\!\! \textstyle\sum\limits_{l=m}^{n-1}\!\!n_l\right)c_{n,\sigma}}\,,\!\!\!\!\!&\text{if } m<n\phantom{\,.}\\[2.5ex]
    \frac{1}{N}&\sum\limits_{\sigma}\expval{c_{m,\sigma}^{\dagger}\exp\left(+i\theta\!\! \textstyle\sum\limits_{l=n}^{m-1}\!\!n_l\right)c_{n,\sigma}}\,,\!\!\!\!\!&\text{if } m>n\phantom{\,.}\\[2.5ex]
    \frac{1}{N}&\sum\limits_{\sigma}\expval{c_{m,\sigma}^{\dagger}c_{m,\sigma}}\,,\!\!\!\!\!&\text{if } m=n\,.
\end{aligned}
\end{cases}
\end{align} 
\newpage
In occupation-dependent-hopping realizations, time-of-flight measurements probe the physical parent particles rather than the intrinsic anyonic operator. For the fermionic realization considered here, the observable physical-fermion momentum distribution
\begin{align}
    n_k^{\mathrm O} =  \dfrac{1}{N}\sum\limits_{\sigma} \expval{c_{k,\sigma}^{\dagger}c_{k,\sigma}} = \dfrac{1}{L}\sum\limits_{m,n=1}^{L} C^\mathrm{O}(m,n) e^{ik(m-n)}
\end{align}
is the Fourier component of the one-body density correlation of spinor fermions
\begin{align}
    C^\mathrm{O}(m,n) = \dfrac{1}{N}\sum\limits_{\sigma}C^{\mathrm{O}}_{\sigma}(m,n) =  \dfrac{1}{N}\sum\limits_{\sigma}\expval{c_{m,\sigma}^{\dagger}c_{n,\sigma}}\,.
\end{align}

\subsubsection{Spin structure factor} 
The structure factor associated with two-body spin correlations is expressed as
\begin{align}
    S(k) =  \dfrac{1}{L}\sum\limits_{m,n=1}^{L} S(m,n) e^{ik(m-n)}\,,
    \label{eqn:diaSSF}
\end{align}
where $S(m,n)$ is the two-body spin correlation. Focusing on the diagonal spin correlations, the two-body spin correlation is defined as \cite{mikkelsen_relation_2023,manmana_su_2011}
\begin{align}
    S(m,n) = \dfrac{1}{N(N-1)} \sum\limits_{\alpha \neq\beta} \left(\expval{n_{m,\alpha}n_{n,\alpha}}-\expval{n_{m,\alpha}n_{n,\beta}}\right)\,.
    \label{eqn:tbdsc}
\end{align}

\subsection{Characterizing the momentum distribution asymmetry}

It is well known that the momentum distribution of anyons breaks the $k\leftrightarrow-k$ symmetry. To characterize such effects,
we primarily use measures of
inversion asymmetry and momentum displacement. The normalized inversion asymmetry or mirror asymmetry,
\begin{align}
\label{eq:mirror_metric}
\mathcal{A}_{\mathrm M}
=
{\textstyle\int\limits_0^\pi \abs{n_k-n_{-k}}\,dk}\;\bigg/\;
{\textstyle\int\limits_0^\pi (n_k+n_{-k})\,dk}\,, 
\end{align}
measures the departure from the inversion-symmetric condition
$n_k=n_{-k}$, with $\mathcal{A}_{\mathrm M}=0$ for a mirror-symmetric
distribution. The mean momentum,
\begin{align}
\label{eq:meank_metric}
\bar{k}
=
{\textstyle\int\limits_{-\pi}^{\pi} k n_k\,dk}\;\bigg/\;
{\textstyle\int\limits_{-\pi}^{\pi} n_k\,dk}\,, 
\end{align}
measures the net displacement of spectral weight. Thus, $\mathcal{A}_{\mathrm M}$ quantifies inversion asymmetry, while $\bar{k}$ characterizes associated net displacement. When $n_k$ in Eqs.~\eqref{eq:mirror_metric} and \eqref{eq:meank_metric} is replaced by $n_k^\mathrm{O}$, we obtain $\mathcal{A}_{\mathrm M}^{\mathrm O}$ and $\bar{k}^{\mathrm O}$, respectively, which characterize the corresponding momentum distribution of the physical fermions. In addition to these measures of asymmetric reconstruction, Appendix~\ref{sec:appendB} examines the total redistribution of spectral weight $R$ and the change in momentum width $\Delta\sigma_k$, which quantify the overall reconstruction of the distribution and its broadening or narrowing, respectively.

\begin{figure*}
\centering
\includegraphics{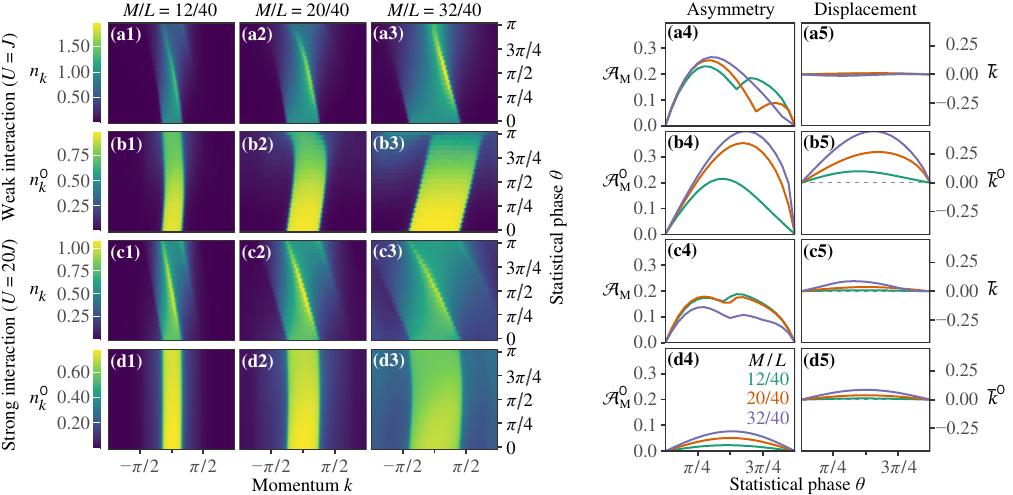}
\caption{\label{fig:anyonsdis}
\textbf{Fractional statistics leaves distinct signatures in intrinsic anyonic and observable physical-fermion momentum distributions.}
Intrinsic anyonic momentum distribution $n_k$ for (a) weak interaction $(U=J)$ and (c) strong interaction $(U=20J)$, and corresponding observable physical-fermion momentum distribution $n_k^{\mathrm O}$ in (b) and (d).
The first three columns show lattice fillings $M/L \in \{12/40,\; 20/40,\; 32/40\}$, respectively, for 21 equally spaced statistical phases $\theta \in \{0,\;0.05\pi,\;\ldots,\;\pi\}$. All panels show $N=2$ components.
Fractional exchange statistics breaks the $k\leftrightarrow-k$ symmetry and redistributes spectral weight; inversion symmetry is restored at fermionic $(\theta=0)$ and pseudobosonic $(\theta=\pi)$ limits.
The fourth and fifth columns show the mirror asymmetry $\mathcal{A}_{\mathrm M}$ [Eq.~\eqref{eq:mirror_metric}] and mean momentum $\bar{k}$ [Eq.~\eqref{eq:meank_metric}], respectively.
For $n_k$, appreciable $\mathcal{A}_{\mathrm M}$ with small
$\bar{k}$ indicates predominantly asymmetric deformation rather
than a uniform momentum shift. 
For $n_k^{\mathrm O}$ at weak interaction, the finite $\mathcal{A}_{\mathrm M}^{\mathrm O}$ and positive $\bar{k}^{\mathrm O}$ show that inversion-symmetry breaking is accompanied by a net momentum displacement.
Through the generalized Jordan--Wigner transformation, $n_k$ explicitly contains the nonlocal statistical string, whereas $n_k^{\mathrm O}$ contains no such string and acquires its $\theta$ dependence indirectly through the many-body state generated by occupation-dependent tunneling.
This distinction becomes pronounced at strong interaction: $n_k$ retains substantial statistics-dependent reconstruction, whereas the corresponding asymmetry and displacement of $n_k^{\mathrm O}$ are strongly suppressed.
Results are obtained using density-matrix renormalization group (DMRG) calculations for system size $L=40$ with open boundary conditions.}
\end{figure*}

\subsection{Two-component anyons}
Spinless anyons provide a useful reference for identifying the role of internal components. Pauli exclusion forbids double occupancy, so the occupation-dependent phase operator does not modify allowed hopping. The Hamiltonian and thus its ground state are $\theta$ independent; any dependence on an observable must arise from its operator representation. For the momentum distribution, this dependence enters solely through the nonlocal Jordan--Wigner string in the anyonic operator \cite{hao_ground-state_2008,hao_ground-state_2009,keilmann_statistically_2011}. 

The situation changes for multicomponent anyons, where particles of different components may occupy the same site. Hopping onto an occupied site then carries the statistical phase, making both the Hamiltonian and the many-body state $\theta$ dependent. The momentum distribution probes the one-body density correlation generated by these hopping processes; in a spinor gas, the same processes also rearrange the internal spin configuration. Charge and spin are therefore intertwined, and the momentum distribution should not retain the form found for spinless anyons \cite{manmana_su_2011,basak_generalized_2023}. The intrinsic anyonic momentum distribution $n_k$ consequently contains two statistical contributions: an explicit contribution from the nonlocal Jordan--Wigner string and an implicit contribution from the $\theta$-dependent many-body state. The observable physical-fermion momentum distribution $n_k^{\mathrm O}$ and the spin structure factor $S(k)$ contain no such string and therefore delineate the state-mediated response. We examine this interplay of $N$ and $\theta$ for two-component anyons in Figs.~\ref{fig:anyonsdis}--\ref{fig:anyonssf}. In the following discussions, we will address the weak- and strong-interaction regimes separately to expose how interaction strength reorganizes the statistical signatures. The complementary momentum-space diagnostics and their connection to the real-space one-body density correlations are developed in Appendices~\ref{sec:appendB}--\ref{sec:appendE}.

\subsubsection*{Weak interaction \texorpdfstring{$(U = J)$}{(U = J)}}

At weak interaction, anyons of different components can readily occupy the same site. Hopping onto such an occupied site carries the statistical phase factor $e^{\pm i\theta}$ and therefore imprints a strong $\theta$ dependence on the many-body state. For fractional exchange statistics, $\theta\in(0,\pi)$, this phase factor is reflected in the complex off-diagonal one-body density correlation $C(m,n)$. For the intrinsic anyonic correlation, the nonlocal Jordan--Wigner string provides an additional statistical phase. \hyperref[fig:anyonsdis]{Figures~\ref{fig:anyonsdis}(a1--3)} show the intrinsic anyonic momentum distribution $n_k$ of two-component anyons at weak $(U=J)$ interaction and lattice fillings $M/L \in \{12/40,20/40,32/40\}$. In the fermionic limit $(\theta=0)$, the two-component fermionic momentum distribution is inversion symmetric and exhibits the characteristic shell structure with strong singularities at $k=\pm k_{\mathrm F}=\pm\pi M/(2L)$. These singularities reflect the characteristic $\pm k_{\mathrm F}$ oscillations of the one-body density correlation; Appendix~\ref{sec:appendC} develops the connection between these real-space oscillations, their power-law decay, and the resulting nonanalytic features in the momentum distribution. This changes dramatically as the statistical phase $\theta$ deviates from $0$. 

Fractional exchange statistics breaks the $k\leftrightarrow-k$ symmetry, redistributes spectral weight, and reconstructs the fermionic shell structure, as also observed for spinless anyons. Using $C(n,m)=C^*(m,n)$, the inversion asymmetry is
\begin{align}
    n_{k}-n_{-k} =  \dfrac{4}{L}\smashoperator[r]{\sum\limits_{m<n=1}^{L}}\Im[C(m,n)] \sin\bm{(}k\left(n-m\right)\bm{)}\,,
    \label{eqn:brokensym}
\end{align}
showing that the inversion asymmetry originates directly from the imaginary part of the one-body density correlation. The general connection between the real and imaginary parts of the one-body density correlation and the even and odd components of $n_k$ is developed in Appendix~\ref{sec:appendC}. Appendix~\ref{sec:appendD} then makes the microscopic origin of the odd component explicit by comparing the intrinsic $N=1$, intrinsic $N=2$, and observable $N=2$ one-body density correlations, thereby separating the operator-string and state-mediated sources of the statistical phase. The resulting momentum-space reconstruction is quantified in Appendix~\ref{sec:appendB} through the redistribution $R$, mirror asymmetry $\mathcal{A}_{\mathrm M}$, mean momentum $\bar{k}$, and width change $\Delta\sigma_k$.

\hyperref[fig:anyonsdis]{Figure~\ref{fig:anyonsdis}(a4)} and \hyperref[fig:anyonsdis]{(a5)} quantify mirror asymmetry $\mathcal{A}_{\mathrm M}$ and mean momentum $\bar{k}$, respectively, distinguishing inversion-symmetry breaking from net spectral-weight displacement. $\mathcal{A}_{\mathrm M}$ is nonmonotonic in $\theta$, becoming largest at intermediate statistical phases and vanishing at both statistical limits. The pronounced $\mathcal{A}_{\mathrm M}$ with small intrinsic $\bar{k}$ indicates the asymmetric reconstruction is dominated by deformation and redistribution of the momentum profile rather than by a rigid momentum shift. The observable physical-fermion momentum distribution $n_k^{\mathrm O}$ in \hyperref[fig:anyonsdis]{Fig.~\ref{fig:anyonsdis}(b1--3)} contains no statistical string and therefore delineates the $\theta$ dependence of the mapped many-body state. The corresponding finite mirror asymmetry $\mathcal{A}_{\mathrm{M}}^{\mathrm{O}}$ in \hyperref[fig:anyonsdis]{Fig.~\ref{fig:anyonsdis}(b4)} and the mean momentum $\bar{k}^{\mathrm{O}}$ in \hyperref[fig:anyonsdis]{Fig.~\ref{fig:anyonsdis}(b5)} therefore originate entirely from this state-mediated response. Importantly, $n_k^{\mathrm O}$ is displaced toward positive momentum [$\bar{k}^{\mathrm{O}} > 0$; \hyperref[fig:anyonsdis]{Fig.~\ref{fig:anyonsdis}(b5)}]. In contrast, $\bar{k} \ll \bar{k}^{\mathrm{O}}$ shows that including the statistical string in the same mapped state counteracts displacement while retaining substantial asymmetry and redistribution. This comparison is consistent with competing string and state effects. Appendix~\ref{sec:appendE} quantifies this interplay through the characteristic wavevectors and decay exponents extracted from the corresponding one-body density correlations.

The emergence of a pronounced $n_k$ peak is reflected as a concentration of spectral weight within a restricted momentum range in \hyperref[fig:anyonsdis]{Fig.~\ref{fig:anyonsdis}(a1--3)}. At fractional exchange statistics, this enhancement occurs at nonzero momentum, providing a visible signature of finite-momentum reconstruction even when the mean momentum remains small. The strong singularities initially at $k=\pm k_{\mathrm F}$ evolve unequally: their positions shift, and one singular feature becomes less pronounced than the other. This asymmetric reconstruction of the fermionic shell structure further characterizes spinor anyons, implying the tendency of the same component anyons to occupy the same state, in contrast to fermionic statistics that prohibits such occupancy.

A second, weaker maximum develops in the intrinsic anyonic momentum distribution, near the second singularity $k \simeq 3k_{\mathrm{F}}$. The emergence of this maximum and its increase with $\theta$ reflect the additional effect of fractional exchange statistics. Finally, the increasing asymmetry of $n_k$ with filling is consistent with the growing probability of configurations in which hopping encounters an occupied site and hence samples the occupation-dependent phase factor. As $\theta$ approaches $\pi$, the pseudobosonic limit, the fermionic singular features become progressively less pronounced while inversion symmetry is restored. However, the pseudobosonic distribution does not necessarily recover the conventional bosonic zero-momentum peak because of the local fermionic exclusion statistics. 

\begin{figure}
\centering
\includegraphics{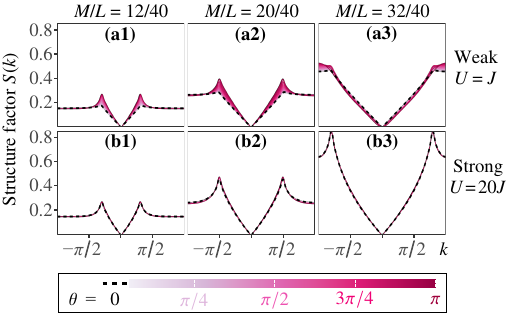}
\caption{\label{fig:anyonssf}\textbf{Spinor anyons retain antiferromagnetic ordering across the full range of fractional exchange statistics.}
Structure factor $S(k)$ obtained from the two-body diagonal spin correlations of two-component anyons [Eq.~\eqref{eqn:tbdsc}]. Columns show lattice fillings $M/L \in \{12/40,\: 20/40,\; 32/40\}$; rows show (a) weak interaction $(U=J)$ and (b) strong interaction $(U=20J)$. The 21 curves span statistical phases $\theta/\pi \in \{0,\;0.05,\;\ldots,\;1\}$: (black dotted) for the fermionic limit $\theta=0$ and progressively darker solid curves for increasing $\theta$. The minimum at $k=0$, approximately linear increase at small $\abs{k}$, and cusps near $k \simeq \pm 2k_{\mathrm F}$ are characteristic of antiferromagnetic spin ordering in Eq.~\eqref{eqn:tbdsc}. The $\theta$ dependence of $S(k)$ progressively weakens with increasing interaction strength and lattice filling. Results are obtained using the density-matrix renormalization group (DMRG) for $L=40$ with open boundary conditions.}
\end{figure}

The contrast between the intrinsic anyonic and observable physical-fermion momentum distributions motivates examining whether an observable constructed from local density operators and containing no explicit anyonic string may exhibit a similarly weakened $\theta$ dependence. The spin structure factor provides precisely this. Figure~\ref{fig:anyonssf}(a) presents the structure factor associated with the two-body diagonal spin correlations. For the varied statistical phases and lattice fillings considered, the spin structure factor demonstrates a similar behavior: minima at $k=0$, linearity for small nonzero values of $k$, maxima at $k=\pm 2k_\mathrm{F}$, followed by decay as $\abs{k}$ increases beyond $2k_\mathrm{F}$. Such behavior is typical of fermionic systems with antiferromagnetic spin ordering. The prohibition of double occupancy of the same spin component as a result of the on-site exclusion statistics realizes the antiferromagnetic ordering. For all fractional exchange statistics, the two-body diagonal spin correlations between neighboring sites remain approximately the same. The correlations beyond the neighboring sites, although small, show a $\theta$ dependence potentially due to boundary effects (open boundary conditions). Consequently, both the slope of the linear region at small $\abs{k}$ and the values of the spin structure factor at its extrema depend weakly on $\theta$. These deviations are suppressed with increasing lattice filling.

Thus, in the weak-interaction regime, fractional exchange statistics acts through two distinct channels. The intrinsic anyonic momentum distribution probes the nonlocal statistical string and the $\theta$-dependent many-body state, whereas the observable physical-fermion momentum distribution contains no statistical string and delineates the state-mediated contribution. The spin structure factor provides a complementary string-free probe of the state-mediated response in spin correlations.

\begin{figure*}
\centering
\includegraphics{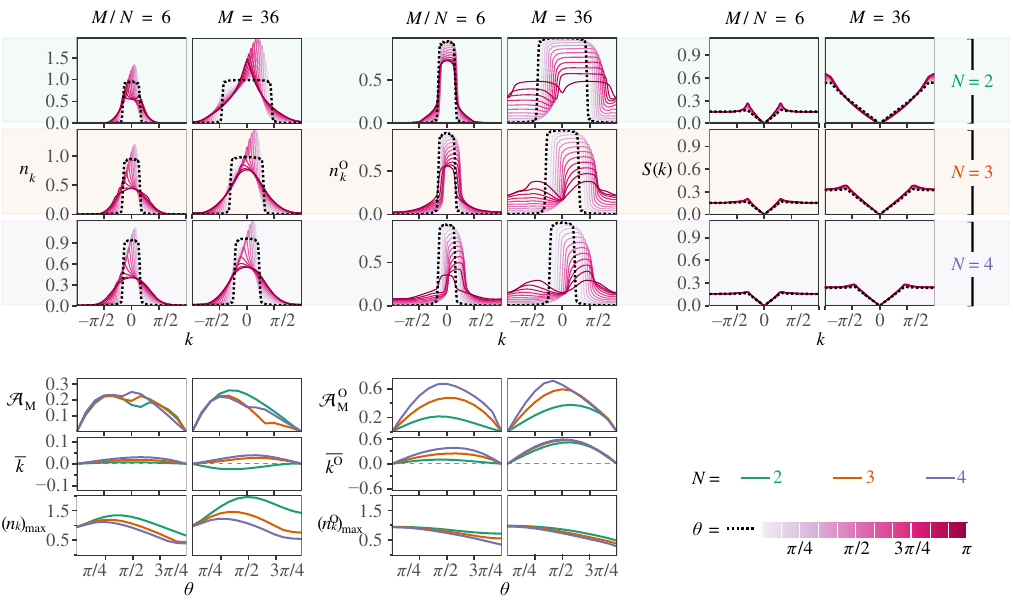}
\caption{\label{fig:anyonsNdis_weak}\textbf{Additional spin components amplify the statistics-dependent momentum reconstruction.}
Shown are the (upper left) intrinsic anyonic momentum distribution $n_k$, (upper middle) observable physical-fermion momentum distribution $n_k^{\mathrm O}$, and (upper right) spin structure factor $S(k)$ at weak interaction $(U=J)$.
Rows show component number $N \in \{2,\;3,\;4\}$; alternating columns show fixed population per component $M/N=6$ and fixed total population $M=36$.
The 13 curves span statistical phases $\theta \in \{0,\;\pi/12,\;\pi/6,\;\pi/4,\;\ldots,\;\pi\}$: (black dotted) for the fermionic limit $\theta=0$ and progressively darker solid curves for increasing $\theta$. 
The lower block shows the (top row) mirror asymmetries $\mathcal{A}_{\mathrm M}$ and $\mathcal{A}_{\mathrm M}^{\mathrm O}$, (middle row) mean momenta $\bar{k}$ and $\bar{k}^{\mathrm O}$, and (bottom row) peak occupations $(n_k)_{\max}\equiv\max_k n_k$ and $(n_k^{\mathrm O})_{\max}\equiv\max_k n_k^{\mathrm O}$.
Increasing $N$ enhances the statistics-dependent redistribution of $n_k$ and, at fixed total population, the mirror asymmetry of $n_k^{\mathrm O}$; $\bar{k}^{\mathrm O}$ shows a weaker dependence on $N$.
In contrast, $S(k)$ remains only weakly dependent on the statistical phase and retains the characteristic structure associated with antiferromagnetic spin ordering.
Results are obtained using DMRG for system size $L=40$ with open boundary conditions.}
\end{figure*}
\begin{figure*}
\centering
\includegraphics{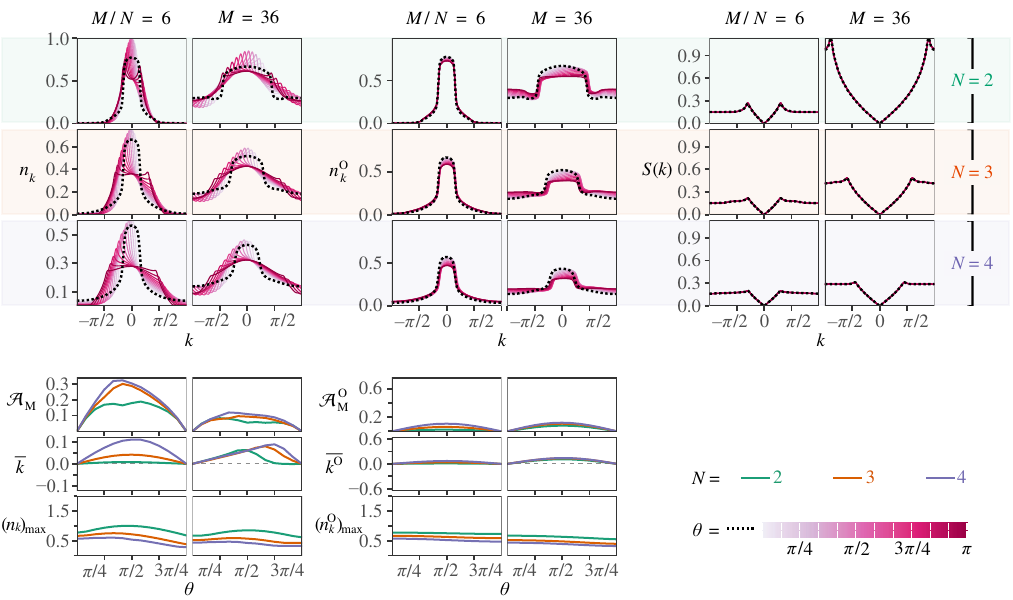}
\caption{\label{fig:anyonsNdis_strong}\textbf{Strong interactions separate intrinsic anyonic signatures from physical-fermion observables'.}
As in Fig.~\ref{fig:anyonsNdis_weak}, except at strong interaction $U=20J$.
The intrinsic anyonic momentum distribution $n_k$ retains substantial dependence on the statistical phase through the nonlocal anyonic string in its defining operator, while the corresponding statistics-dependent signatures in the observable physical-fermion momentum distribution $n_k^{\mathrm O}$ are strongly suppressed.
This suppression is reflected in the small values of both $\mathcal{A}_{\mathrm M}^{\mathrm O}$ and $\bar{k}^{\mathrm O}$ across $N$.
The spin structure factor $S(k)$ likewise remains nearly insensitive to $\theta$ and retains the characteristic antiferromagnetic form.}
\end{figure*}

\subsubsection*{Strong interaction \texorpdfstring{$(U = 20J)$}{(U = 20J)}}

At strong interaction, double occupancy is suppressed and the low-energy spin and charge sectors decouple. To leading order, the charge sector is identical to that of the corresponding spinless system, and therefore contributes the same charge correlations to the momentum distribution. However, in the multicomponent case, the spin sector remains as an effective spin chain (for a detailed discussion, see Ref.~\cite{basak_anyon2}). Each hopping process in the charge sector is accompanied by a corresponding rearrangement of the spin configuration. The momentum distribution is thus obtained by weighting the spinless charge correlations by the overlap associated with the spin rearrangement. As double occupancy is suppressed, the state-mediated statistical-phase dependence is strongly reduced.

\hyperref[fig:anyonsdis]{Figures~\ref{fig:anyonsdis}(c1--3)} show that $n_k$ remains sensitive to fractional exchange statistics.  
In addition to the distribution broadening, the secondary peak becomes stronger as a result of strong intercomponent interaction. Increasing the lattice filling, or equivalently the number of anyons, predictably increases the distribution width. In contrast to weak interaction, a finite $\bar{k}$ is also observed to emerge and becomes increasingly pronounced with filling. As was the case in weak interaction, restoration of inversion symmetry at $\theta=\pi$ does not imply restoration of a conventional bosonic momentum distribution; the pseudobosonic limit instead reflects the interplay of deformed statistics, filling, and interaction strength.

$n_k^{\mathrm O}$ shows a markedly different behavior at strong interaction. The curves become nearly independent of $\theta$ \hyperref[fig:anyonsdis]{[Fig.~\ref{fig:anyonsdis}(d1--3)]}, and the associated momentum-space diagnostics are strongly suppressed (Appendix~\ref{sec:appendB}). Consistently, the one-body density correlations analyzed in Appendix~\ref{sec:appendE} retain characteristic wavevectors close to the fermionic values $\pm k_{\mathrm F}$ and exhibit only weak $\theta$ dependence in their decay exponents. The state-mediated statistical response is therefore strongly reduced as anticipated.

In contrast, $n_k$ retains a substantial $\theta$ dependence at the same interaction strength. Even when the state-mediated contribution is suppressed, its asymmetry and redistribution survive as a consequence of the nonlocal statistical string. This aspect is also directly visible in the imaginary part of the one-body density correlation in Appendix~\ref{sec:appendD}.

The spin structure factor retains the antiferromagnetic form seen at weak interaction, but the curves for different $\theta$ nearly overlap, particularly at larger filling [Fig.~\ref{fig:anyonssf}]. Since the spin correlations are constructed from local density operators and contain no explicit anyonic string, their approximate $\theta$ independence at strong interaction mirrors that of the observable physical-fermion momentum distribution. The microscopic origin of this hierarchy between intrinsically anyonic and state-mediated statistical signatures is developed in Ref.~\cite{basak_anyon2}.

\subsection{\texorpdfstring{$N$}{N}-component anyons}
We now extend our study beyond two components, considering $N \in \{2,\;3,\;4\}$. The ground state is examined with statistical phase $\theta \in [0,\;\pi]$ using the Anyon--Hubbard Hamiltonian for system size $L=40$ and open boundary conditions, fixing either population per component ($M/N=6$) or total population ($M = 36$). Figures~\ref{fig:anyonsNdis_weak} and \ref{fig:anyonsNdis_strong} show the corresponding momentum distributions and spin structure factors at weak and strong interactions, respectively. In the fermionic limit, the intrinsic anyonic momentum distribution retains a shell structure with singular features at $k=\pm k_{\mathrm F}=\pm(\pi/N)(M/L)$. Fractional exchange statistics continuously reconstructs this shell structure as $\theta$ is increased, while increasing $N$ generally redistributes spectral weight toward larger $\abs{k}$ and broadens the momentum distribution.

\subsubsection*{Weak interaction \texorpdfstring{$(U = J)$}{(U = J)}}
Fractional exchange statistics further reconstructs the momentum distributions as $N$ increases [Fig.~\ref{fig:anyonsNdis_weak}]. Here, particles of different components can occupy the same site, so increasing $N$ enlarges the accessible intercomponent configuration space, including configurations with larger local occupation $n_j$. For a given occupation $n_0$, the number of possible component configurations grows as $\binom{N}{n_0}$. Consequently, at weak interaction, a larger number of many-body configurations and hopping pathways can involve an already occupied site and therefore sample the occupation-dependent phase factor $e^{i \theta n_j}$. Increasing $N$ thus allows the system to sample and accumulate a broader set of phase factors $(e^{i \theta n_0})$. This provides a microscopic origin for the enhancement, with increasing $N$, of $\theta$-dependent spectral-weight redistribution and broadening of both the intrinsic anyonic and observable physical-fermion momentum distributions. Appendix~\ref{sec:appendB} shows that this trend persists at both fixed population per component and fixed total population, demonstrating it is not merely a consequence of increasing the particle number.

Appendix~\ref{sec:appendC} provides a useful real-space picture of this reconstruction. The characteristic wavevectors and power-law decay exponents of the oscillations in the one-body density correlation determine, respectively, the positions and sharpness of the nonanalytic features in the momentum distributions, both of which depend on $N$ and $\theta$. Considering the contribution from the nonlocal statistical string, the fermionic wavevectors $\pm k_{\mathrm{F}}$, with $k_{\mathrm{F}} = \pi M/(NL)$, are shifted approximately by $N (\theta/\pi) k_{\mathrm{F}}$ to $k_+$ and $k_-$, while their separation remains as $2k_{\mathrm F}$. Fractional exchange statistics also produces unequal decay exponents, with $\alpha_{-}>\alpha_{+}$, so that the singularity at $k_{-}$ becomes smoother than that at $k_{+}$. These expressions provide an approximate picture of the explicit string dependence; the additional changes seen in the physical-fermion distribution arise from the $\theta$- and $N$-dependent many-body state.

The peak height provides another view of this reconstruction. For the intrinsic distribution, $(n_k)_{\max}$ varies nonmonotonically with $\theta$, increasing at small to intermediate fractional exchange statistics before decreasing toward the pseudobosonic limit. At fractional exchange statistics, the peak height also generally decreases as $N$ is increased, for both fixed $M/N$ and fixed $M$, consistent with spectral weight being spread over a broader range of momenta. The observable distribution behaves differently: $(n_k^{\mathrm O})_{\max}$ decreases more steadily with $\theta$, with the reduction becoming stronger as $N$ increases.

The weak-interaction spin structure factor retains a central minimum, an approximately linear small-$\abs{k}$ region, and finite-momentum cusps for all component numbers [Fig.~\ref{fig:anyonsNdis_weak}]. The minimum at $k=0$ confirms the antiferromagnetic ordering. Increasing $N$ modifies the magnitude and shape of these features, while a visible $\theta$ dependence persists most clearly at lower filling. Increasing the filling reduces this dependence.

Thus, in the weak-interaction regime, both momentum and spin sectors respond to fractional exchange statistics, but the momentum distribution exhibits more pronounced reconstruction. The corresponding diagnostics in Appendix~\ref{sec:appendB} show this component dependence is expressed most systematically through the enhanced total redistribution and broadening, while asymmetry and displacement depend more sensitively on whether total or per-component population is held fixed.

\subsubsection*{Strong interaction \texorpdfstring{$(U = 20J)$}{(U = 20J)}}

At strong interaction, suppression of multiple occupancy produces a pronounced fermionization of the momentum distribution [Fig.~\ref{fig:anyonsNdis_strong}]. Spectral weight is transferred away from small momentum and spread over a broader range of $k$, producing pronounced finite-momentum and shell-like features. This behavior becomes increasingly pronounced as $N$ is increased. Fractional exchange statistics further reconstructs this fermionized profile: for $0<\theta<\pi$, the distribution becomes asymmetric, with spectral weight transferred differently on either side of $k=0$, while the finite-momentum peaks are also reshaped. In the pseudobosonic limit, inversion symmetry is restored, but strong interaction prevents a return to the smooth weak-interaction profile; instead, increasing $N$ produces a broad, deformed shell-like distribution.

The intrinsic peak height remains nonmonotonic in $\theta$, with a maximum at fractional exchange statistics, but decreases as $N$ is increased. Thus, fractional exchange statistics can enhance a finite-momentum feature within the fermionized distribution, while increasing $N$ spreads the spectral weight more broadly over the shell-like profile. In contrast, $(n_k^{\mathrm O})_{\max}$ shows only a weak dependence on $\theta$ at strong interaction and decreases primarily with increasing $N$, consistent with the suppression of the state-mediated statistical response.

The observable physical-fermion momentum distribution and the spin structure factor show a much weaker dependence on the statistical phase [Fig.~\ref{fig:anyonsNdis_strong}]. Although increasing $N$ continues to modify their overall structure, the curves for different $\theta$ become nearly overlapping, particularly at larger filling. Thus, strong interaction enhances the fermionization associated with increasing $N$, while simultaneously suppressing the $\theta$ dependence carried by the many-body state. The intrinsic anyonic momentum distribution remains an exception because the nonlocal anyonic operator continues to retain the statistical phase.

This distinction is also reflected in the diagnostics of Appendix~\ref{sec:appendB}: the observable momentum response is strongly suppressed, whereas the intrinsic distribution retains component-dependent redistribution and asymmetry even when its width change becomes small. This weak $\theta$ dependence of the spin sector at strong interaction is better examined microscopically using the generalized effective spin-chain (GESC) formalism developed for strongly interacting anyons in Ref.~\cite{basak_anyon2}.

\section{\label{sec:timeev}Transport dynamics}
We next examine how fractional exchange statistics and interaction shape transport, focusing on dipole oscillations following a displacement of the harmonic trap and the expansion of a localized impurity upon release in a two-component gas. These protocols provide complementary probes: dipole oscillations reveal the predominantly charge-sector collective response, while impurity release probes statistics-dependent, spin-resolved transport through the host gas.
Here, we study transport in a two-component anyon gas trapped in a 1D optical lattice using the time-dependent variational principle (TDVP) method (developed with input from Ref.~\cite{mendl_pytenet_2018}).

\subsection{Dipole oscillations}

We study the dipole dynamics of a spin-balanced gas of two-component anyons following a sudden displacement of the trap center. The system comprises $M=10$ particles in a 1D optical lattice of size $L=20$ with an additional weak harmonic trap, with equal populations $M^{\uparrow}=M^{\downarrow}=M/2$. The resulting dipole oscillations probe the collective motion of the many-body gas in response to an applied force and, therefore, provide a time-domain measure of its global transport response. Importantly, the center-of-mass operator is constructed entirely from the local density and is invariant under the anyon--fermion Jordan--Wigner transformation. It therefore contains no explicit statistical phase. Any $\theta$ dependence of the dipole dynamics must instead originate from the many-body state as it evolves under the statistics-dependent Hamiltonian. This distinction becomes important for multicomponent anyons: for spinless fermion-based anyons, the statistical phase drops out of the hopping Hamiltonian, whereas for $N>1$ it remains in the Hamiltonian and can affect the dynamics. Consequently, the oscillation frequency, amplitude, and damping provide direct measures of how fractional exchange statistics modifies collective transport through the many-body dynamics.

\subsubsection*{System}
The Hamiltonian in the presence of an external harmonic trap is defined, assuming lattice spacing as $1$, as
\begin{align}
    H = &-J\smashoperator{\sum\limits_{\langle i,j\rangle,\sigma}}(a_{i,\sigma}^{\dagger}a_{j,\sigma}+\text{H.c.})\notag\\&+\sum\limits_i \left( U n_{i}^{\uparrow}n_{i}^{\downarrow} + \Omega \left[x_i-\delta \Theta(t)\right]^2 n_i \right) \,,
\end{align}
where $x_i = (L+1)/2-i$ denotes the position of site $i$ from the center of the chain, $n_i =\sum_{\sigma}n_{i,\sigma}$ is the local density, $\delta$ is the displacement of the trap center, and $\Theta(t)$ is the unit step function ($= 1$ for $t>0$, and $0$ otherwise). Aside from the tunneling $J$ and on-site interaction $U$, the Hamiltonian includes an additional harmonic trap $\Omega$.

\subsubsection*{Evolution}
Prior to the evolution, the trap center is shifted. Its sudden displacement results in dipole oscillations, characterized by the oscillations in the center of mass $[x_{\mathrm{com}} = (\sum_i x_{i} \langle n_{i} \rangle)/M]$. \textls[-1]{The $x_{\mathrm{com}}$ oscillations arise from the periodic redistribution of the density about the displaced trap center.} In spinor fermions or bosons, the nature of the oscillations depends on the displacement of the trap center, harmonic trap strength, on-site interaction, and density; the response ranges from underdamped or undamped $x_{\mathrm{com}}$ oscillations about the trap center to overdamped relaxations \cite{huo_dipolar_2012,montangero_dipole_2009,rigol_collective_2005}. Consideration of anyons with $N$ internal components enables the use of the statistical phase as a variable to tune the nature of these oscillations.

\begin{figure*}
\centering
\includegraphics[width=5.75in]{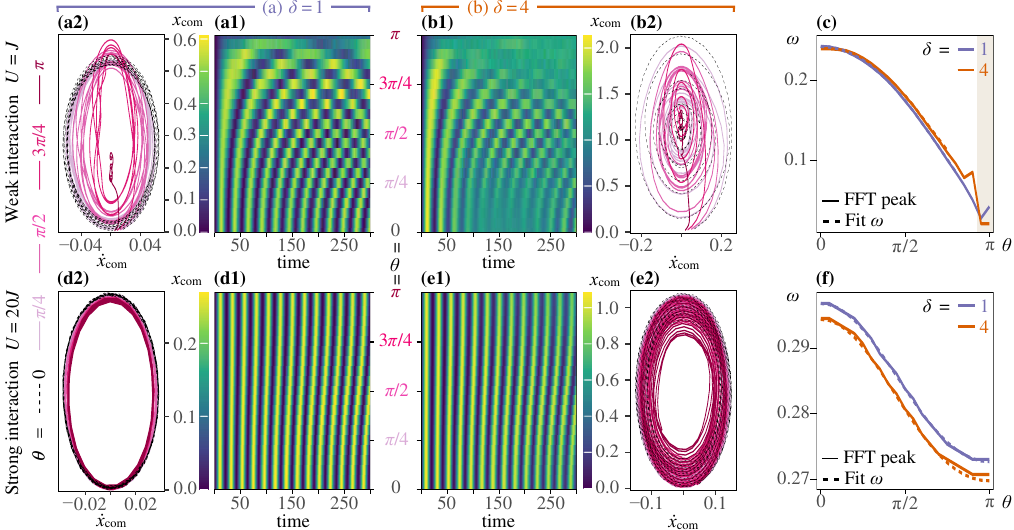}
\caption{\label{fig:anyonscom}\textbf{Fractional exchange statistics controls the frequency, amplitude, and damping of dipole oscillations.}
Center-of-mass $x_{\mathrm{com}}$ dynamics following a sudden displacement of a weak harmonic trap $(\Omega = 5 \times 10^{-3}J)$. 
Upper and lower rows show weak interaction $(U=J)$ and strong interaction $(U=20J)$, respectively.
Panels (1) show the $x_{\mathrm{com}}$ response following trap-center displacements $\delta=1$ and $4$, with all 21 statistical phases $\theta/\pi \in \{0,\;0.05,\;\ldots,\;1\}$.
Panels (2) show corresponding phase-space trajectories (\textit{n.b.}~velocity plotted on x-axis to share the $x_{\mathrm{com}}$ y-axis) for selected $\theta \in \{0,\;\pi/4,\;\pi/2,\;3\pi/4,\;\pi\}$.
Panels (c) and (f) compare frequencies from validated fits to the empirical damped-cosine model $[x_{\mathrm{com}}(t)=Ae^{-\beta t}\cos(\omega t)+B]$ (dashed) with dominant Fourier peak frequencies (solid); purple and orange denote $\delta=1$ and $4$, respectively.
Fit curves are interrupted where reliability checks fail.
Shading marks $\theta$ intervals where at least one displacement has a boundary-limited estimate or fewer than two fitted cycles.
At weak interaction, varying $\theta$ strongly modifies the oscillation frequency and damping and can produce nonoscillatory relaxation. At strong interaction, persistent oscillatory bands and nearly overlapping phase-space trajectories indicate weaker damping and reduced sensitivity to statistics. Results are obtained using the time-dependent variational principle (TDVP) method for $L=20$, with a spin-balanced population, $M=10$, and open boundary conditions. Time is expressed in units of $J^{-1}$.}	
\end{figure*}

Figure~\ref{fig:anyonscom} shows the center-of-mass dynamics for a weak harmonic trap $(\Omega=5\times10^{-3}J)$, with small $(\delta=1)$ and large $(\delta=4)$ displacements. The time--statistics heatmaps (panels 1) show $x_{\mathrm{com}}(t,\theta)$, and the rightmost panels (c,f) show the extracted dominant oscillation frequency. The corresponding phase-space trajectories (panels 2) provide a complementary perspective on the system dynamics by plotting $\dot{x}_{\mathrm{com}}$ against $x_{\mathrm{com}}$. Their radial extent reflects the oscillation amplitude, whereas their inward contraction quantifies the decay of coherent motion. Consequently, underdamped oscillations manifest as repeated, weakly decaying loops, while strongly damped or nonoscillatory relaxation is characterized by a trajectory that converges to the stationary point without looping cycles.

\subsubsection*{Weak interaction \texorpdfstring{$(U = J)$}{(U = J)}}

The initial density is concentrated near the harmonic-trap center, so a substantial fraction of the cloud participates in the subsequent motion. For a small displacement, the fermionic gas exhibits clearly resolved underdamped oscillations [Fig.~\ref{fig:anyonscom}(a1)]. Increasing $\theta$ progressively reduces both the frequency and amplitude of the dipole oscillations and enhances their damping, eventually producing strongly damped or nonoscillatory relaxation over the time interval shown. The extracted frequency in Fig.~\ref{fig:anyonscom}(c) correspondingly decreases strongly with $\theta$, demonstrating a pronounced softening of the dipole mode, while the $x_{\mathrm{com}}(t,\theta)$ heatmaps show the simultaneous suppression of the oscillatory motion. Toward the pseudobosonic limit, the oscillations become both very slow and strongly damped, with the center-of-mass motion becoming nearly arrested. This evolution is also evident in the phase-space trajectories in Fig.~\ref{fig:anyonscom}(a2): increasing $\theta$ reduces their extent and accelerates their contraction toward the stationary point, with the strongly damped trajectories completing progressively fewer rotations. At weak interaction, particles of different components can still occupy the same site, so the displaced spinor gas repeatedly samples the occupation-dependent statistical phase during tunneling. The resulting slowing and damping therefore reflect the statistics-dependent many-body dynamics of the spinor gas, in which the spatial and internal-component degrees of freedom remain coupled.

The large-displacement $\delta=4$ response is strongly nonmonotonic in $\theta$. This is seen directly in the heatmap [Fig.~\ref{fig:anyonscom}(b1)] from the contrast of the late-time oscillations: starting from the fermionic limit, the amplitude initially increases with $\theta$ and reaches a maximum around $\theta/\pi\simeq0.5$--$0.6$, indicating reduced damping and more persistent dipole oscillations at intermediate fractional exchange statistics. The phase-space trajectories in Fig.~\ref{fig:anyonscom}(b2) show the same behavior. Their broad extent reflects the large oscillation amplitude, while their clear separation for different $\theta$ values highlights the strong sensitivity of the dynamics to fractional exchange statistics. As the cloud oscillates, the trajectories contract inward toward the stationary point associated with the displaced trap, thereby directly depicting the damping of the collective motion and its approach toward the new trap center. This contraction is slowest at intermediate values of $\theta$, consistent with the enhanced persistence seen in the heatmap, and becomes more rapid again toward the pseudobosonic limit. At larger $\theta$, the oscillatory amplitude decreases rapidly, while the dominant frequency continues to decrease \hyperref[fig:anyonscom]{[Fig.~\ref{fig:anyonscom}(c)]}. Close to the pseudobosonic limit, the center-of-mass motion becomes both very slow and strongly suppressed. Thus, for a large quench, fractional exchange statistics produces a crossover from underdamping at intermediate $\theta$ to overdamping as $\theta\rightarrow\pi$.

Notably, the characteristic frequencies for $\delta=1$ and $4$ follow nearly the same statistical-phase dependence despite their very different amplitude dynamics. This indicates that the statistics-induced reduction of the dipole frequency is largely displacement independent, while the oscillation damping depends much more strongly on the quench amplitude.

\subsubsection*{Strong interaction \texorpdfstring{$(U = 20J)$}{(U = 20J)}}

The initial density is more uniformly distributed across the occupied region, and the center-of-mass oscillation amplitudes are correspondingly smaller than for $U=J$. However, persistent oscillatory bands and repeated phase-space orbits with weak contraction indicate underdamped motion for both $\delta=1$ and $4$ over the time interval shown [Fig.~\ref{fig:anyonscom} (bottom row)]. The slowly decaying oscillatory envelope with weak variation with $\theta$ demonstrates the reduced sensitivity to fractional exchange statistics. Increasing $\theta$ still reduces the oscillation amplitude and frequency. However, these changes are substantially smaller than at weak interaction. Thus, strong interaction suppresses the $x_{\mathrm{com}}$ response magnitude while preserving sustained oscillations. The charge sector remains $\theta$ independent to leading order, while the spin sector retains weak statistical dependence. Residual statistical effects beyond this leading-order limit can accumulate during evolution and contribute to the small $\theta$ dependence of $x_{\mathrm{com}}$. In the strong-interaction regime, this reduced statistical sensitivity parallels that of the observable physical-fermion momentum distribution and spin correlations in the ground state.

\subsection{Impurity release dynamics}
We study the release dynamics of a single impurity immersed in a gas of hard-core anyons using a spinor gas. The system comprises two-component anyons trapped in a weak harmonic trap in addition to the lattice with system size $L=21$ and number of particles $M=10$. To recreate the physics of an impurity in hard-core anyons, $M^{\uparrow}=M-1$ spin-up ($\uparrow$) anyons form the hard-core host gas and a single spin-down ($\downarrow$) anyon acts as the impurity. The impurity is coupled to the gas via intercomponent interaction.

\subsubsection*{System}
The Hamiltonian governing the system is given by
\begin{align}
    H =&-J\smashoperator{\sum\limits_{\langle i,j\rangle,\sigma}}(a_{i,\sigma}^{\dagger}a_{j,\sigma}+\text{H.c.})\nonumber\\&+\sum\limits_i \left( U n_{i}^{\uparrow}n_{i}^{\downarrow} + \Omega x^{2}_{i} n_i\right) +h_{z} \Theta(-t)\sigma^{z}_{i_c}\,,
\end{align}
where $\sigma^{z}_{i_c}$ is the local spin-polarization operator coupled to the pinning field $h_z=4J$. It preferentially locates the impurity ($\downarrow$) at the center of the system [$i_c = (L+1)/2$]. At $t=0$, the pinning field is quenched to zero. Besides the tunneling $J$, on-site interaction $U$, and the local external magnetic field, the Hamiltonian includes a weak harmonic trap $(\Omega = 5 \times 10^{-3} J)$.

\begin{figure}
\centering
\includegraphics{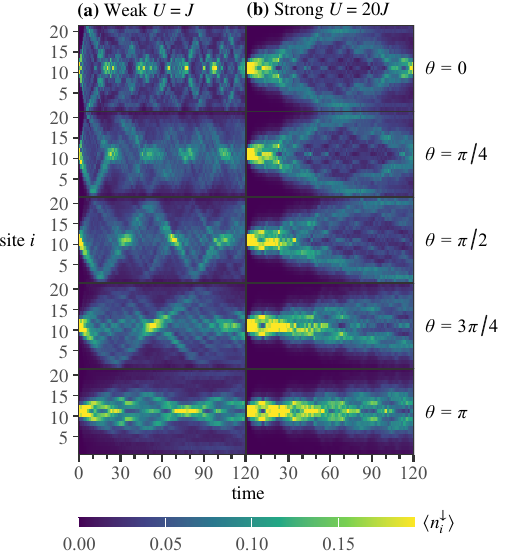}
\caption{\label{fig:irir}\textbf{Fractional exchange statistics suppresses and skews impurity expansion.}
Evolution of a single spin-down impurity density $\langle{n_i^\downarrow}\rangle$ in a gas of spin-up anyons for (a) weak $(U=J)$ and (b) strong $(U=20J)$ interactions. Local pinning field $h_z=4J$ is removed at $t=0$.
Rows correspond to statistical phases $\theta \in \{0,\;\pi/4,\;\pi/2,\;3\pi/4,\;\pi\}$.
Intermediate statistical phases produce inversion-asymmetric propagation at both interaction strengths.
At weak interaction, increasing $\theta$ progressively suppresses the spatial expansion and momentum-space breathing (Appendix~\ref{sec:appendG}).
At strong interaction, the fronts propagate more slowly, with the expansion being slowest in the pseudobosonic limit.
Results are obtained for a weak harmonic trap $(\Omega=5\times10^{-3}J)$ using TDVP for $L=21$, $M=10$, $M^\uparrow=M-1$, and open boundary conditions.
Colors saturate at the 99th percentile.
Time is expressed in units of $J^{-1}$.
}
\end{figure}

\subsubsection*{Evolution}
We consider the impurity ($\downarrow$) initially localized at the system center and immersed in a gas of hard-core anyons ($\uparrow$). Prior to the evolution, the local external magnetic field is turned off. The quench dynamics of the impurity released from the center of the system is examined. Figure~\ref{fig:irir} presents the evolution of the impurity population at weak interaction ($U=J$) and at strong interaction ($U=20J$) for statistical phases $\theta \in \{0,\;\pi/4,\;\pi/2,\;3\pi/4,\;\pi\}$. 

\subsubsection*{Weak interaction \texorpdfstring{$(U = J)$}{(U = J)}}
A fermionic impurity released from the center develops two counterpropagating fronts that reach the system edges and refocus [Fig.~\ref{fig:irir}(a)]. Fractional exchange statistics makes propagation inversion asymmetric, while increasing $\theta$ progressively slows expansion and concentrates the impurity closer to the center. At $\theta=\pi$, inversion symmetry is restored, but the impurity remains substantially more localized than in the fermionic limit. The asymmetry originates from the interplay of occupation-dependent tunneling and on-site interaction in the mapped Hamiltonian. Each nearest-neighbor hop acquires a statistical phase set by the local occupation, so different hopping paths/histories accumulate different phases and interfere differently with their inversion-related counterparts. This unequal interference is an outcome of on-site interaction simultaneously distinguishing the many-body configurations encountered along these paths. This produces asymmetric impurity expansion; Appendix~\ref{sec:appendF} illustrates the mechanism at short times. This path-interference mechanism connects the present dynamics to earlier work on asymmetric anyonic expansion and correlated-hopping transport \cite{liu_anyon_2018,greschner_probing_2018,kwan_realization_2024}. Recent two-component calculations with synthetic gauge flux provide a further comparison \cite{ChenHuangZhangZhang2026}; here the asymmetry is obtained for a released impurity in a many-particle host with separately conserved components and no additional synthetic flux.

The weak-interaction momentum-space dynamics is presented in Appendix~\ref{sec:appendG}. In the fermionic limit, the impurity exhibits pronounced momentum-space breathing. Increasing $\theta$ progressively suppresses both the amplitude and frequency of this breathing, providing a momentum-space counterpart of the statistics-induced suppression of expansion in real space.

\subsubsection*{Strong interaction \texorpdfstring{$(U = 20J)$}{(U = 20J)}}
The impurity expansion is more strongly suppressed from the outset, and its dynamics depends only weakly on $\theta$ [Fig.~\ref{fig:irir}(b)]. The inversion asymmetry visible at fractional exchange statistics for $U=J$ is substantially reduced, and the impurity remains localized near the central region for all statistical phases considered. Strong interaction suppresses real double occupancy and thereby reduces the influence of occupation-dependent tunneling on the density evolution. The impurity dynamics therefore follows the same hierarchy found in the ground-state physical-fermion and spin observables: state-mediated statistical signatures become much weaker even though intrinsically anyonic observables can remain strongly dependent on the fractional exchange statistics.

\section{\label{sec:summary}Summary and Outlook}
Our work establishes how fractional exchange statistics fundamentally reshapes the ground-state and transport properties of multicomponent anyons in one-dimensional optical lattices. Analyzing the multicomponent Anyon--Hubbard model across varying statistical phases, interaction strengths, lattice fillings, and number of internal components, we reveal behaviors that depart qualitatively from those of conventional spinor fermions and bosons. Ultimately, exchange statistics emerges as a versatile, independent control knob for tuning both equilibrium correlations and nonequilibrium quantum transport. \hyperref[tab:summary]{Table~\ref{tab:summary}} summarizes the main findings, their microscopic origins, and the resulting opportunities for controlling correlations and transport through fractional exchange statistics.

Fractional exchange statistics can strongly reshape the momentum distribution. Increasing the statistical phase reconstructs the symmetric fermionic shell structure, generating asymmetry, finite-momentum peaks, and modified singularities. In the pseudobosonic limit, the conventional zero-momentum peak can instead give way to finite-momentum peaks, forming a quasi-fermionic shell structure highly sensitive to filling and interaction strength. Extension to multicomponent anyons further reveals the interplay of fractional exchange statistics and multicomponent fermionization: increasing component number redistributes spectral weight toward larger momenta, with a strongly statistical-phase-dependent response. This highlights how internal degrees of freedom provide an additional control knob to reshape statistics-driven charge correlations.
\onecolumngrid
\vfill

\begin{table}[!b]
\caption{\label{tab:summary}
\textbf{Fractional exchange statistics as a control knob for tuning momentum correlations and quantum transport.}
Main findings in this work and their microscopic origins, highlighting measurable signatures and routes to controlling momentum distributions, spin correlations, and transport through the statistical phase $\theta$, interaction strength $U$, and number of internal components $N$.}

\setlength{\tabcolsep}{3pt}
\renewcommand{\arraystretch}{1.75}

\begin{ruledtabular}
\begin{tabular}{
>{\raggedright\arraybackslash}p{0.16\textwidth}
>{\raggedright\arraybackslash}p{0.28\textwidth}
>{\raggedright\arraybackslash}p{0.26\textwidth}
>{\raggedright\arraybackslash}p{0.24\textwidth}
}
\textbf{Observable / regime}
&
\textbf{Principal finding(s)}
&
\textbf{Physical origin}
&
\textbf{Experimental signatures and opportunities for control}
\\
\hline

Momentum reconstruction: weak interaction
&
Intrinsic distribution develops pronounced asymmetry despite small mean momentum. Physical-fermion distribution exhibits displacement.
&
The explicit statistical string and the statistics-dependent many-body state produce competing changes in the one-body density correlation.
&
Tune the structure of the intrinsic momentum distribution; distinguish string- and state-mediated signatures.
\\[7pt]

Momentum and spin structure: strong interaction
&
Intrinsic momentum distribution retains $\theta$ dependence. The physical-fermion distribution and spin structure factor become nearly $\theta$ independent. Antiferromagnetic ordering persists.
&
Suppressed multiple occupancy weakens the state-mediated response, while the intrinsic one-body density correlation retains its nonlocal statistical string.
&
Reshape intrinsic momentum correlations while approximately preserving spin structure, providing selective control over observables of the same gas.
\\[7pt]

Increasing number of components $N$
&
Enhances statistics-induced redistribution and broadening at weak $U$, including at fixed total population. Asymmetry and displacement depend on the population protocol.
&
Additional components modify the accessible intercomponent configurations, spin rearrangements, and relative weights of the momentum branches.
&
Use internal-state multiplicity with $\theta$ to control momentum profiles at fixed density, extending available control beyond interaction strength and filling.
\\[7pt]

Collective dipole motion
&
Weak $U$: $\theta$ tunes frequency, amplitude, and damping; large-quench oscillations at intermediate $\theta$ and overdamping toward $\pi$. Strong $U$: weak $\theta$ sensitivity.
&
Occupation-dependent tunneling modifies collective motion; strong repulsion suppresses its influence on the leading charge dynamics.
&
Tune collective oscillation frequencies and relaxation at fixed $U$ and trap parameters, with $U$ controlling $\theta$ sensitivity.
\\[7pt]

Impurity propagation and momentum breathing
&
$\theta$ produces asymmetric propagation and suppresses impurity expansion and momentum-space breathing. Strong interaction slows the fronts, while residual  statistical dependence remains.
&
Statistical phases modify interference between hopping pathways; spin-resolved propagation also probes internal rearrangements and finite-interaction processes.
&
Tune impurity mobility, directional asymmetry, and breathing dynamics. Compare impurity and collective motion to probe spin and charge dynamics.
\\
\end{tabular}
\end{ruledtabular}
\end{table}

\newpage
\twocolumngrid
The spin sector responds distinctly from the charge sector. The spin structure factor retains clear signatures of antiferromagnetic ordering, yet its statistical-phase dependence rapidly weakens as on-site interaction strengthens. Consequently, strongly interacting spinor anyons can simultaneously harbor highly statistics-dependent intrinsic momentum distributions and nearly statistics-independent spin correlations, reflecting the persistence of nonlocal statistical strings in the former even as the state-mediated statistical response weakens.

Fractional exchange statistics also provides direct control over macroscopic transport. Tuning the statistical phase dictates the amplitude, frequency, and damping of dipole oscillations; capable of driving crossover from persistent oscillations to strongly damped relaxation. For a released impurity, fractional exchange statistics yields distinct inversion-asymmetric propagation, while approaching the pseudobosonic limit heavily suppresses spatial propagation and progressively quenches momentum-space breathing. In the strong-interaction regime, real-space density dynamics remains largely immune to the statistical phase across these protocols, even as momentum-space observables retain robust statistical signatures.

Collectively, these results establish the statistical phase as a powerful tunable parameter for controlling fermionization, collective relaxation, impurity transport, and momentum-space breathing within the same microscopic system. Furthermore, they uncover a physical theme: statistical effects dominate observables that explicitly retain the anyonic exchange phase, whereas strong interaction can systematically wash out these effects in purely density-based or collective dynamics.


\paragraph*{Future avenues.}Looking ahead, this framework naturally opens compelling avenues for the exploration of statistics-driven quantum phase transitions, dynamical fermionization, and long-time thermalization following interaction or statistics quenches. Introducing component-dependent statistical phases, spin-dependent interactions, and higher internal symmetries promises to unveil qualitatively new spin--charge behavior. Ultimately, extending this work to more general driving protocols and experimentally accessible preparation and detection schemes should firmly establish fractional exchange statistics as a tangible, programmable resource for the design and control of correlated one-dimensional quantum matter.
\vspace*{-1em}
\begin{acknowledgments}
\vspace*{-1em}
This study was supported by the National Science Foundation under Grant No.~PHY-2513089 and The Welch Foundation under Grant No.~C-1669.
SB acknowledges support by the National Science Foundation through Grant No.~PHY-2409311.
XWG acknowledges support from the National Natural Science Foundation of China (NSFC) Key Grants Nos.~U25D8013, 92365202, 12465001 and the Innovation Program for Quantum Science and Technology Grant Nos.~2021ZD0302000 and 2023ZD0300404.
This work was supported in part by the Big-Data Private-Cloud Research Cyberinfrastructure MRI-award funded by NSF under grant CNS-1338099 and by Rice University's Center for Research Computing (CRC). 
Some of the computing for this project was performed at the OU Supercomputing Center for Education \& Research (OSCER) at the University of Oklahoma (OU).
\end{acknowledgments}

\appendix
\let\addtocontents\origaddtocontents
\begin{center}
\PRLsep
\small\textbf{\\APPENDICES\vspace{-2em}}
\end{center}
\begingroup
\makeatletter
\renewcommand{\tocname}{}
\renewcommand{\baselinestretch}{1}\selectfont
\setlength{\parskip}{6pt}
\def\l@f@section{\addpenalty{\@secpenalty}}
\vspace{-4em}
\tableofcontents
\par
\makeatother
\endgroup
\section{\label{sec:appendA}Anyon--fermion mapping}
Use the Jordan--Wigner transformation to map the anyons to nonlocal fermionic operators:
\begin{align}
    a_{j,\sigma} = \exp\left(-i\theta \sum\limits_{l<j}n_l\right)c_{j,\sigma}\,.
\end{align}
\subsection*{Commutation rules}
The mapping ensures the nonlocal fermionic operators satisfy the generalized commutation rules. We verify this by substituting the mapping into $a_{j,\alpha}a_{k,\beta}^{\dagger}$ and using the fermionic anticommutation relations:
{\small
\begin{align*}
    &a_{j,\alpha}a_{k,\beta}^{\dagger}  \\& = \exp\left(-i\theta \sum\limits_{l<j}n_l\right)c_{j,\alpha}c_{k,\beta}^{\dagger}\exp\left(+i\theta \sum\limits_{l<k}n_l\right)\\ & =
\begin{cases}
    \begin{aligned}-c_{k,\beta}^{\dagger}&\exp\left(+i\theta \smashoperator{\sum\limits_{l<k,\neq j}}n_l\right) \\&\times\exp\left(-i\theta \sum\limits_{l<j}n_l\right)c_{j,\alpha}\exp\left(+i\theta n_j\right)\end{aligned}\,,& \text{if } j<k\phantom{\,.}\\~\\
    \delta_{\alpha\beta}- c_{j,\beta}^{\dagger}c_{j,\alpha}=\delta_{\alpha\beta}- a_{j,\beta}^{\dagger}a_{j,\alpha}\,,& \text{if } j=k\phantom{\,.}\\~\\
    \begin{aligned}-\exp\left(-i\theta n_k\right)c_{k,\beta}^{\dagger}&\exp\left(+i\theta \sum\limits_{l<k}n_l\right) \\&\times\exp\left(-i\theta \smashoperator{\sum\limits_{l<j,\neq k}}n_l\right)c_{j,\alpha}\end{aligned}\,,& \text{if } j>k\,.
\end{cases}
\end{align*}}
For $j<k$, commuting $c_{j,\alpha}$ through the local phase factor:
{\small\begin{align*}
    &a_{j,\alpha}a_{k,\beta}^{\dagger} (j<k) \\&  = -c_{k,\beta}^{\dagger}\exp\left(+i\theta \smashoperator{\sum\limits_{l<k,\neq j}}n_l\right) \exp\left(-i\theta \sum\limits_{l<j}n_l\right)\exp\left[i\theta \left(n_j+1\right)\right]c_{j,\alpha}
    \\ & = -\exp\left(+i\theta\right)c_{k,\beta}^{\dagger}\exp\left(+i\theta \sum\limits_{l<k}n_l\right) \exp\left(-i\theta \sum\limits_{l<j}n_l\right)c_{j,\alpha}
    \\ &  = -\exp\left(+i\theta \right) a_{k,\beta}^{\dagger} a_{j,\alpha} = -\exp\left[-i\theta \sgn(j-k)\right] a_{k,\beta}^{\dagger} a_{j,\alpha} \,.
\end{align*}}
Similarly, for $j>k$, commuting the local phase factor through $c_{k,\beta}^{\dagger}$:
{\small
\begin{align*}
      &a_{j,\alpha}a_{k,\beta}^{\dagger} (j>k) \\&   = -c_{k,\beta}^{\dagger}\exp\left[-i\theta \left(n_k+1\right)\right]\exp\left(+i\theta \sum\limits_{l<k}n_l\right) \exp\left(-i\theta \smashoperator{\sum\limits_{l<j,\neq k}}n_l\right)c_{j,\alpha}
    \\ &  = -\exp\left(-i\theta\right)c_{k,\beta}^{\dagger}\exp\left(+i\theta \sum\limits_{l<k}n_l\right) \exp\left(-i\theta \sum\limits_{l<j}n_l\right)c_{j,\alpha}
    \\ &  = -\exp\left(-i\theta \right) a_{k,\beta}^{\dagger} a_{j,\alpha} = -\exp\left[-i\theta \sgn(j-k)\right] a_{k,\beta}^{\dagger} a_{j,\alpha} \,.
\end{align*}}
Combining these results with the on-site anticommutation relation, we obtain
\begin{align}
a_{j,\alpha}a_{k,\beta}^{\dagger} +e^{-i\theta \sgn(j-k)} a_{k,\beta}^{\dagger}a_{j,\alpha} = \delta_{j,k}\delta_{\alpha\beta}\,.
\end{align}
The exchange relation between two annihilation operators follows
{\small
\begin{align*}
     &a_{j,\alpha}a_{k,\beta}  \\& = \exp\left(-i\theta \sum\limits_{l<j}n_l\right)c_{j,\alpha}\exp\left(-i\theta \sum\limits_{l<k}n_l\right)c_{k,\beta}\\ & =
\begin{cases}
    \begin{aligned}-&\exp\left(-i\theta \smashoperator{\sum\limits_{l<k,\neq j}}n_l\right)c_{k,\beta}\\&\times\exp\left(-i\theta \sum\limits_{l<j}n_l\right)c_{j,\alpha}\exp\left(-i\theta n_j\right)\end{aligned}\,,& \text{if } j<k\\~\\
    -\exp\left(-2i\theta\sum\limits_{l<j}n_l\right)c_{j,\beta}c_{j,\alpha}\,,& \text{if } j=k\\~\\
    \begin{aligned}-\exp\left(-i\theta \sum\limits_{l<k}n_l\right)&\exp\left(-i\theta n_k\right)c_{k,\beta}\\&\times\exp\left(-i\theta \smashoperator{\sum\limits_{l<j,\neq k}}n_l\right)c_{j,\alpha}\end{aligned}\,,& \text{if } j>k
\end{cases}
\end{align*}}
For $j<k$, we find
{\small
\begin{align*}
\begin{split}
    & a_{j,\alpha}a_{k,\beta} (j<k) \\&  = -\exp\left(-i\theta \smashoperator{\sum\limits_{l<k,\neq j}}n_l\right)c_{k,\beta}\exp\left(-i\theta \sum\limits_{l<j}n_l\right)\exp\left[-i\theta \left(n_j+1\right)\right]c_{j,\alpha}
    \\ & = -\exp\left(-i\theta \right)\exp\left(-i\theta \sum\limits_{l<k}n_l\right)c_{k,\beta}\exp\left(-i\theta \sum\limits_{l<j}n_l\right)c_{j,\alpha}
    \\ &  = -\exp\left(-i\theta \right) a_{k,\beta} a_{j,\alpha} = -\exp\left[i\theta \sgn(j-k)\right] a_{k,\beta} a_{j,\alpha} \,,
\end{split}
\end{align*}}
For $j>k$, we find
{\small
\begin{align*}
\begin{split}
    & a_{j,\alpha}a_{k,\beta} (j>k) \\&  = -\exp\left(-i\theta \sum\limits_{l<k}n_l\right)c_{k,\beta}\exp\left[-i\theta \left(n_k-1\right)\right]\exp\left(-i\theta \smashoperator{\sum\limits_{l<j,\neq k}}n_l\right)c_{j,\alpha}
    \\ &  = -\exp\left(+i\theta\right)\exp\left(-i\theta \sum\limits_{l<k}n_l\right)c_{k,\beta}\exp\left(-i\theta \sum\limits_{l<j}n_l\right)c_{j,\alpha}
    \\ &  = -\exp\left(+i\theta \right) a_{k,\beta} a_{j,\alpha} = -\exp\left[i\theta \sgn(j-k)\right] a_{k,\beta} a_{j,\alpha} \,,
\end{split}
\end{align*}
}
In combination with the on-site relation, these expressions give
\begin{align}
a_{j,\alpha}a_{k,\beta} + e^{+i\theta \sgn(j-k)} a_{k,\beta}a_{j,\alpha}=0\,.
\end{align}

\subsection*{Fermi--Hubbard Hamiltonian} To obtain the mapped Fermi--Hubbard Hamiltonian, apply the Jordan--Wigner transformation to the nearest-neighbor hopping terms:
\begin{align*}
\begin{split}
      a_{i,\alpha}^{\dagger}a_{i+1,\alpha} & = c_{i,\alpha}^{\dagger}\exp\left(+i\theta \sum\limits_{l<i}n_l\right) \exp\left(-i\theta \sum\limits_{l<i+1}n_l\right)c_{i+1,\alpha}
    \\ &  = c_{i,\alpha}^{\dagger} \exp\left(-i\theta n_i\right)c_{i+1,\alpha} \,,
\end{split}
\end{align*}
The Hermitian-conjugate term becomes
\begin{align*}
\begin{split}
      a_{i+1,\alpha}^{\dagger}a_{i,\alpha} & = c_{i+1,\alpha}^{\dagger}\exp\left(+i\theta \sum\limits_{l<i+1}n_l\right) \exp\left(-i\theta \sum\limits_{l<i}n_l\right)c_{i,\alpha}
    \\ &  = c_{i+1,\alpha}^{\dagger} \exp\left(+i\theta n_i\right)c_{i,\alpha} \,,
\end{split}
\end{align*}
Unchanged under this mapping, the local densities are
\begin{align*}
\begin{split}
      n_{i,\alpha} & = a_{i,\alpha}^{\dagger}a_{i,\alpha}
    \\ &  = c_{i,\alpha}^{\dagger}\exp\left(+i\theta \sum\limits_{l<i}n_l\right) \exp\left(-i\theta \sum\limits_{l<i}n_l\right)c_{i,\alpha}
    \\ &  = c_{i,\alpha}^{\dagger}c_{i,\alpha} = n_{i,\alpha} \,,
\end{split}
\end{align*}
Thus, the mapping introduces occupation-dependent tunneling phases while preserving on-site interactions:
\begin{align}
\begin{aligned}H_{\mathrm{F}} = &-J\sum\limits_{j,\alpha} \left(c_{j,\alpha}^{\dagger}e^{-i\theta n_{j}}c_{j+1,\alpha} + c_{j+1,\alpha}^{\dagger}e^{+i\theta n_{j}} c_{j,\alpha}\right) \\&+\sum\limits_{i,\alpha<\beta} U_{\alpha\beta}n_{i,\alpha}n_{i,\beta}\,.\end{aligned}
\end{align}

\section{\label{sec:appendB}\texorpdfstring{$N$}{N}-component momentum reconstruction}
\begin{figure}
\centering
\includegraphics{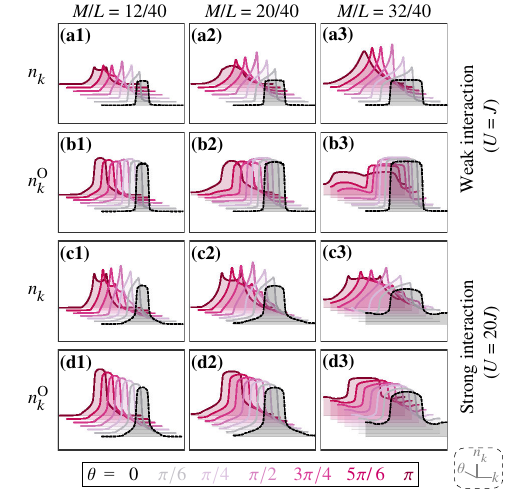}
\caption{\label{fig:slices_anyonsdis}\textbf{Representative momentum-distribution cuts at selected statistical phases.}
Intrinsic anyonic momentum distributions $n_k$ [(a) and (c)] and observable physical-fermion momentum distributions $n_k^{\mathrm O}$ [(b) and (d)] corresponding to the full statistical-phase maps in Fig.~\ref{fig:anyonsdis}.
Panels (a) and (b) show weak interaction $(U=J)$, while panels (c) and (d) show strong interaction $(U=20J)$.
Columns correspond to lattice fillings $M/L=12/40$, $20/40$, and $32/40$.
Each panel shows the selected statistical phases $\theta \in \{0,\;0.15\pi\approx\pi/6,\;\pi/4,\;\pi/2,\;3\pi/4,\;0.85\pi\approx5\pi/6,\;\pi\}$.
The fermionic limit $\theta=0$ is shown as a black dotted curve, while progressively darker solid curves represent increasing $\theta$.
Curves are vertically offset for clarity, as illustrated by the inset.
System parameters and numerical details are as in Fig.~\ref{fig:anyonsdis}.}
\end{figure}

\begin{figure*}
\centering
\includegraphics[width=6in]{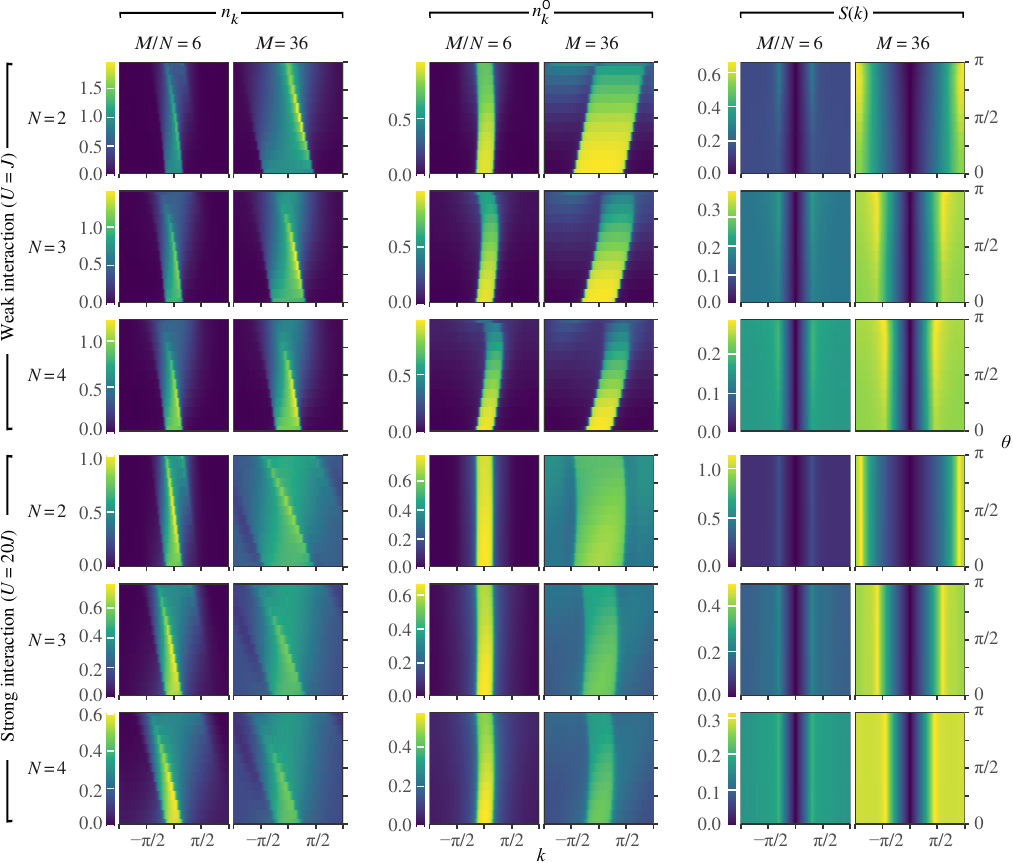}
\caption{\label{fig:heatmaps_anyonsNdis}\textbf{Heatmap representation of the momentum distributions and spin structure factors in Figs.~\ref{fig:anyonsNdis_weak} and \ref{fig:anyonsNdis_strong}.} Shown are the (left) intrinsic anyonic momentum distribution $n_k$, (middle) observable physical-fermion momentum distribution $n_k^{\mathrm O}$, and (right) spin structure factor $S(k)$ versus momentum $k$ and statistical phase $\theta$.
The upper and lower blocks correspond to weak $(U=J)$ and strong $(U=20J)$ interaction, respectively.
Within each block, rows show $N=2$, $3$, and $4$ components and alternating columns correspond to fixed population per component $M/N=6$ and fixed total population $M=36$.
The heatmaps highlight how the component number $N$ shapes the $\theta$-dependent momentum-space reconstruction, modifying the redistribution of spectral weight, asymmetry, and prominence of finite-momentum features.
Results are obtained using density-matrix renormalization group (DMRG) calculations for system size $L=40$ with open boundary conditions.}
\end{figure*}

In this section, we characterize the reconstruction of the momentum distributions seen in Figs.~\ref{fig:anyonsdis}, ~\ref{fig:anyonsNdis_weak}, and ~\ref{fig:anyonsNdis_strong} as functions of component number $N$ and statistical phase $\theta$. Representative momentum cuts at selected statistical phases of the distributions seen in Fig.~\ref{fig:anyonsdis} are shown in Fig.~\ref{fig:slices_anyonsdis}, illustrating finite-momentum reconstruction induced by fractional exchange statistics. Complementing these cuts, Fig.~\ref{fig:heatmaps_anyonsNdis} presents heatmaps of the momentum distributions in Figs.~\ref{fig:anyonsNdis_weak} and \ref{fig:anyonsNdis_strong}, highlighting their evolution with statistical phase $\theta$ for different component numbers $N$. We use four complementary diagnostic measures to quantify different aspects of this reconstruction in the intrinsic anyonic momentum distribution $n_k$ and the observable physical-fermion momentum distribution $n_k^{\mathrm O}$ with $N$ and $\theta$. The normalized momentum distribution
\begin{align}
    p_{\theta}(k) = \dfrac{n_k(\theta)}{\int n_k(\theta)\,dk}\,,
\end{align}
such that $\int p_{\theta}(k) dk =1$, is used to obtain the four measures. These are the redistribution $R$, mirror asymmetry $\mathcal{A}_{\mathrm{M}}$, mean momentum $\bar{k}$, and momentum-width change $\Delta\sigma_k$, given by
\begin{align}
  R(\theta)
  &=
  \frac{1}{2}\int \abs{p_{\theta}(k)-p_{0}(k)}\,dk\,,
  \label{eq:R}
  \\
  \mathcal{A}_{\mathrm{M}}(\theta)
  &=
  \frac{
    \displaystyle\int_{k>0}
    \abs{p_{\theta}(k)-p_{\theta}(-k)}\,dk
  }{
    \displaystyle\int_{k>0}
    \left[p_{\theta}(k)+p_{\theta}(-k)\right]\,dk
  }\,,
  \label{eq:AM}
  \\
  \bar{k}_\theta
  &=
  \int k p_{\theta}(k)\,dk\,,
  \label{eq:mean}
  \\\Delta\sigma_k(\theta)
  &=
  \sigma_k(\theta)-\sigma_k(0)\,, \nonumber \\&\text{ with }~ \sigma_k(\theta) = \sqrt{\int
    \left(k-\bar{k}_\theta\right)^2
    p_{\theta}(k)\,dk}\,.
  \label{eq:width}
\end{align}

\begin{turnpage}
\begin{figure*}
\centering
\includegraphics[width=\linewidth]{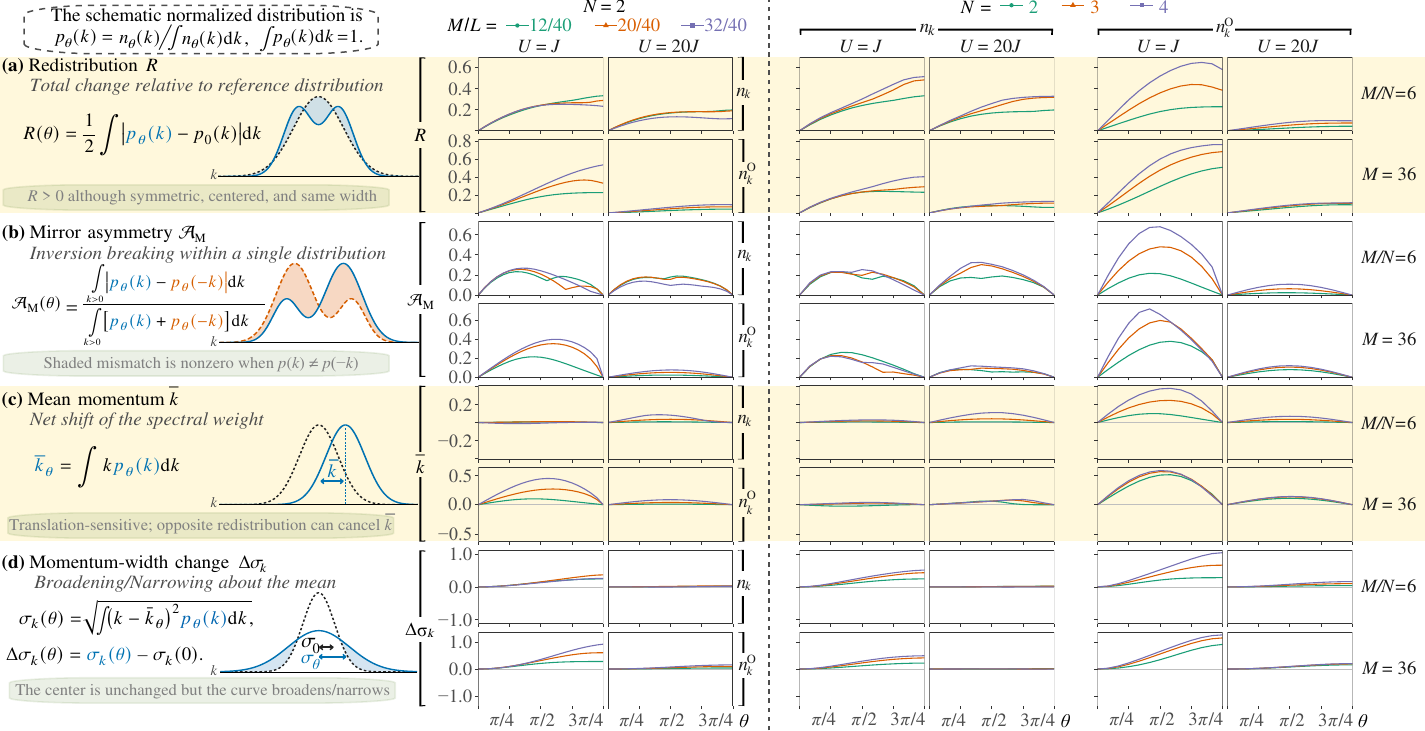}
\caption{\label{fig:allmetrics}\textbf{Complementary diagnostics separate momentum reconstruction from inversion breaking, displacement, and broadening.}
The schematics on the left define the four measures obtained from the normalized momentum distribution
$p_\theta(k)=n_\theta(k)/\int n_\theta(k)\,dk$:
(a) the redistribution relative to the fermionic reference,
$R(\theta)=\frac{1}{2}\int\abs{p_\theta(k)-p_0(k)}\,dk$;
(b) the mirror asymmetry $\mathcal{A}_{\mathrm M}$, which measures inversion breaking within a single distribution;
(c) the mean momentum $\bar{k}$, which measures the signed displacement of spectral weight; and
(d) the width change
$\Delta\sigma_k(\theta)=\sigma_k(\theta)-\sigma_k(0)$, which distinguishes broadening from narrowing about the mean.
The central block presents these diagnostics for the two-component system with $L=40$, fillings $M/L \in \{12/40,\;20/40,\;32/40\}$, and interactions $U=J$ and $U=20J$. Within each metric, the upper row shows the intrinsic anyonic distribution $n_k$ and the lower row shows the physical-fermion distribution $n_k^{\mathrm O}$. 
The right block compares $N \in \{2,\;3,\;4\}$ components at the same interaction strengths, with pairwise columns for $n_k$ and $n_k^{\mathrm O}$ and pairwise rows for fixed population per component $M/N=6$ $(M=6N \in \{12,18,24\})$ and fixed total population $M=36$. 
The curves connect 21 $\theta/\pi \in \{0,\;0.05,\;\ldots,\;1\}$ in the central block and 13 $\theta \in \{0,\;\pi/12,\;\ldots,\;\pi\}$, without fitting or smoothing.
Together, the four measures demonstrate that substantial symmetric momentum-space reconstruction persists even when the mirror asymmetry and mean displacement vanish, and that filling, component number, and interaction strength affect the magnitude and form of the reconstruction differently.
}
\end{figure*}
\end{turnpage}

$R$ measures the total reconstruction relative to the fermionic distribution at $\theta=0$.  It detects both symmetric and asymmetric changes, but does not specify the weight rearrangement. $\mathcal{A}_{\mathrm{M}}$ captures the inversion symmetry breaking, $p_{\theta}(k) \neq p_{\theta}(-k)$, but vanishes for a symmetric distribution even when the distribution deviates from the $\theta=0$ distribution. $\bar{k}$ represents the net shift of the momentum distribution away from $k =0$. Equivalently put, it is the weighted center of the momentum distribution. As rearrangements on opposite sides of $k=0$ can balance each other, a small $\bar{k}$ does not necessarily imply that the momentum distribution has changed weakly. Finally, $\Delta\sigma_k$ distinguishes broadening from narrowing. It can miss changes in the shape of the distribution that leave its overall width unchanged. 

All four diagnostics are calculated by numerically integrating the momentum distributions on their original momentum grids. Their physical meanings are illustrated schematically in Fig.~\ref{fig:allmetrics}(a--d). Columns 2 and 3 show the diagnostics for the intrinsic and observable physical-fermion momentum distributions for $N=2$, interaction strengths $U=J$ and $U=20J$, and lattice fillings $M/L \in \{12/40,\;20/40,\;32/40\}$. Columns 4--7 extend this analysis to $N \in \{2,\;3,\;4\}$, for the same interaction strengths, comparing fixed population per component, $M/N = 6$, with fixed total population, $M=36$.

For $N=2$, at weak interaction and low filling, the redistribution $R$ generally increases with $\theta$. At higher lattice fillings, however, its $\theta$ dependence becomes nonmonotonic. The mirror asymmetry $\mathcal{A}_{\mathrm{M}}$ is nonzero only at fractional exchange statistics, and likewise varies nonmonotonically with $\theta$. For the observable distribution, $\mathcal{A}_{\mathrm{M}}^{\mathrm{O}}$ exhibits a single maximum at intermediate $\theta$, whereas for the intrinsic distribution a single maximum evolves into two local maxima as the filling is decreased. At the same time, the mean momentum $\bar{k}_\theta$ remains comparatively small for the intrinsic distribution but becomes substantial for the observable distribution. This contrast is consistent with the two sources of $\theta$ dependence contributing differently to the two momentum distributions, with partial cancellation possible in the intrinsic distribution. Positive $\Delta\sigma_k(\theta)$ further indicates both distributions generally broaden as $\theta$ is increased. At strong interaction, most of these responses are strongly suppressed, with the main exception being the mirror asymmetry of the intrinsic distribution.

Extending the comparison to $N = 2, 3, 4$, increasing the number of components at weak interaction generally enhances the statistics-induced redistribution $R$ and the broadening $\Delta\sigma_k(\theta)$ of both intrinsic anyonic and observable physical-fermion momentum distributions. Importantly, this trend persists whether $M/N =6$ or $M=36$, showing it is not solely caused by the increase in particle number. The mirror asymmetry and mean momentum of the observable distribution also generally increase with $N$. For the intrinsic distribution at weak interaction, the component dependence of the asymmetry depends on how the population is held fixed. For $M/N=6$, the asymmetry generally increases with $N$, whereas at fixed $M=36$ it decreases with $N$. The intrinsic mean momentum remains comparatively small in both cases, despite the stronger redistribution and broadening. At strong interaction, the distinction between the two momentum distributions becomes more pronounced. All four diagnostics of the observable distribution are strongly suppressed. The intrinsic distribution, in contrast, retains a component-dependent redistribution, asymmetry, and mean momentum, while its width change becomes very small. Thus, adding components does not simply amplify a single statistics-dependent effect, but changes how the momentum-space reconstruction is divided among redistribution, inversion asymmetry, displacement, and broadening.

Taken together, the diagnostics show that $N$ and $\theta$ play distinct but coupled roles in shaping the momentum distributions. Increasing $N$ most systematically strengthens the overall redistribution and broadening, particularly at weak interaction, while $\theta$ controls the character of this reconstruction: asymmetry and displacement are strongest at intermediate $\theta$ and vanish again at the symmetric endpoints. Importantly, the recovery of inversion symmetry toward $\theta = \pi$ does not imply a return to the fermionic distribution, since substantial redistribution and broadening can remain. At strong interaction, these statistics-dependent signatures are strongly suppressed in $n_k^{\mathrm O}$, whereas $n_k$ retains an appreciable and component-dependent reconstruction. The following sections connect these momentum-space diagnostics to the real-space one-body density correlation, first by identifying the oscillation scales and algebraic decay expected in limiting cases and then by separating the state- and operator-induced sources of the $\theta$ dependence.

\section{\label{sec:appendC} Asymptotic one-body correlations and fits}
To complement the diagnostics in Appendix~\ref{sec:appendB}, here we revisit the asymptotic form of the one-body density correlation. The purpose is not to obtain an exact correlation function for $N$-component anyons in a 1D optical lattice, but to extract the characteristic wavevectors and the exponents associated with the power-law decay in limiting cases. This motivates the functional form used for the numerical analysis below.

Define the spatially averaged one-body density correlation
\begin{align}
C_{\mathrm{avg}}(r)
=
\frac{1}{L-r}\sum_{m=1}^{L-r}
C(m,m+r)\,.
\end{align}
The momentum distribution can then be written as
\begin{align}
n_k
=
C_{\mathrm{avg}}(0) +\frac{2}{L}\sum_{r=1}^{L-1} (L-r)
\bigg\{&\Re [C_{\mathrm{avg}}(r)\cos(kr)] \nonumber
\\ +
&\Im[C_{\mathrm{avg}}(r)\sin(kr)]
\bigg\}\,.
\label{eq:obdm_to_nk}
\end{align}
The real part of the correlation therefore determines the even contribution to the momentum distribution, whereas its imaginary part generates the odd contribution,
\begin{align}
n_k-n_{-k}
=
\frac{4}{L}\sum_{r=1}^{L-1}
(L-r) \Im[C_{\mathrm{avg}}(r)\sin(kr)]\,.
\label{eq:obdm_odd}
\end{align}
\textls[-10]{The emergence of an imaginary part of the one-body density correlation at fractional exchange statistics is thus a direct real-space signature of the inversion asymmetry quantified by $\mathcal A_{\mathrm M}$.}

For noninteracting spinless fermions,
\begin{align}
C_{\mathrm{avg}}(r)
=
\frac{\sin(k_{\mathrm F}r)}{\pi r}
=
\frac{e^{+ik_{\mathrm F}r}-e^{-ik_{\mathrm F}r}}
{2i\pi r}\,.
\label{eq:free_spinless_obdm}
\end{align}
The correlation thus contains equal contributions with characteristic wavevectors $\pm k_{\mathrm F}$, both decaying as $r^{-1}$. The wavevectors determine the positions of the nonanalytic features in momentum space, while the power-law exponent determines their character and sharpness. Equation~\eqref{eq:free_spinless_obdm} yields the familiar zero-temperature Fermi distribution with discontinuities at $\pm k_{\mathrm F}$.

For spinless anyons, the generalized Jordan--Wigner transformation introduces the nonlocal occupation-dependent statistical string
\begin{align}
C_{\mathrm{avg}}(r)
=
\frac{1}{L-r} \sum_m \expval{c_m^\dagger
e^{-i\theta\mathcal N_{m,r}}
c_{m+r}}
\,,
\end{align}
where $\mathcal N_{m,r}=\sum_{j=m}^{m+r-1}n_j $. Writing
\begin{align}
\mathcal N_{m,r}
=
\rho r+\delta\mathcal N_{m,r}\,,
\end{align}
with $\rho=\frac{M}{L}$, separates the mean particle number from its fluctuations. Neglecting $\delta\mathcal N_{m,r}$ gives
\begin{align}
C_{\mathrm{avg}}(r)
\approx
e^{-i\kappa k_{\mathrm F}r}
\frac{\sin(k_{\mathrm F}r)}{\pi r}\,,
\end{align}
where the reduced statistical phase $\kappa=\theta/\pi$. The mean-density contribution thus shifts the two characteristic wavevectors to
\begin{align}
k_+&=(1-\kappa)k_{\mathrm F}\,,
\\k_-&=-(1+\kappa)k_{\mathrm F}\,.
\label{eq:spinless_kpm}
\end{align}
The two remain separated by $2k_{\mathrm F}$, but their midpoint is displaced by $-\kappa k_{\mathrm F}$. Thus, the mean part of the statistical-phase-dependent string produces a momentum shift without changing the $1/r$ envelope. The particle-number fluctuations modify this result. Within the low-energy Luttinger-liquid description, retaining the fluctuations gives
\begin{align}
C_{\mathrm{avg}}(r)
\sim
A_+\frac{e^{ik_+r}}{r^{\alpha_+}}
+
A_-\frac{e^{ik_-r}}{r^{\alpha_-}}\,,
\label{eq:spinless_anyon_asymp}
\end{align}
with
\begin{align}
\alpha_\pm
=
1\mp\kappa+\frac{\kappa^2}{2}\,.
\label{eq:spinless_anyon_alpha}
\end{align}
This structure is consistent with the established Luttinger-liquid description and exact asymptotic results for 1D anyonic fluids~\cite{calabrese_correlation_2007,patu_correlation_2007,santachiara_one-particle_2008}.

For $0<\kappa<1$, $\alpha_+<1$ while $\alpha_->1$. The $k_+$ contribution therefore becomes increasingly dominant at long distances, sharpening the corresponding nonanalytic momentum-space feature, while the $k_-$ feature becomes progressively smoother. Fractional exchange statistics therefore produces both a displacement of the characteristic momentum scales and an unequal scaling of the two branches, providing a simple microscopic origin for the skewing of the momentum distribution.

To incorporate $N$ internal components, we switch back to fermions. For noninteracting, balanced $N$-component fermions the components are independent, and each satisfies  
 \begin{align}
C_{\mathrm{avg},\sigma}(r)
=
\frac{\sin(k_{\mathrm F}r)}{\pi r}\,,
\end{align}
where, now in the presence of $N$ internal components, $k_{\mathrm F} = \pi \rho/N$.  The component-averaged correlation ($C_{\mathrm{avg}}(r)$) retains the same $r^{-1}$ decay and the same pair of characteristic wavevectors $\pm k_{\mathrm F}$. Increasing $N$ does not introduce a new exponent; at fixed total population it instead changes $k_{\mathrm F}$ through the reduced population per component ($M/N$).

In the presence of repulsive interactions, the correlation retains its $\pm k_{\mathrm F}$ oscillations to leading order but decays with exponent~\cite{Kawakami1993Exponents,yamamoto_universal_2023}
\begin{align}
\alpha_{\mathrm F}
=
1+\frac{(K_{\mathrm c}-1)^2}{2NK_{\mathrm c}}
=
\frac{N-1}{N}
+
\frac{1}{2N}
\left(
K_{\mathrm c}+\frac1{K_{\mathrm c}}
\right)\,,
\label{eq:sun_fermion_alpha}
\end{align}
where the first contribution originates from the $N-1$ spin modes and the second from the charge sector. As interaction strength increases from $0 \to \infty$, $K_{\mathrm c}$ goes from $1$ to $1/N$.

The extension to spinor anyons is less direct. Unlike the spinless case, even at $U=0$ the generalized Jordan--Wigner mapping generates a $\theta$-dependent Hamiltonian and thus a $\theta$-dependent mapped fermionic many-body state. Thus, in addition to the explicit $\theta$ dependence of the operator string, there is another dependence from the state. Related multicomponent anyon models have likewise shown statistics-dependent ground-state structure and correlations~\cite{santos_quantum_2012,patu_correlation_2019}. Even so, a string-only reference at fixed mapped state---which varies the anyonic string while holding the corresponding same-$N$ fermionic state fixed---is useful for isolating one source of statistical dependence in the momentum distribution. Thus, the statistical phase only enters through the nonlocal string of the operator. This shall give us an asymptotic form originating from just the string, 
\begin{align}
C^{\mathrm{str}}_{\mathrm{avg}}(r)
\sim
A_+
\frac{e^{ik_+r}}{r^{\alpha_+}}
+
A_-
\frac{e^{ik_-r}}{r^{\alpha_-}}\,,
\label{eq:spinor_anyon_string_form}
\end{align}
\begin{subequations}
\label{eq:spinor_anyon_ansatz}
\begin{flalign}
\text{with~}k_{+}
&\approx (1-N\kappa)k_{\mathrm F}\,,
\label{eq:spinor_anyon_kplus}
\\
k_{-}
&\approx -(1+N\kappa)k_{\mathrm F}\,,
\label{eq:spinor_anyon_kminus}
\\
\text{and~}\alpha_\pm
&\approx
\frac{N-1}{N}
+
\frac{1}{2N}
\left[
K_{\mathrm c}(1\mp N\kappa)^2
+
\frac{1}{K_{\mathrm c}}
\right]\,.&&
\label{eq:spinor_anyon_ansatz_alpha}
\end{flalign}
\end{subequations}
Equations~\eqref{eq:spinor_anyon_ansatz} are not exact predictions for the spinor-anyon lattice model, but serve as reference forms for the numerical analysis.  The approximate form above captures the contribution of the nonlocal statistics-dependent string. Comparing the $N=1$ intrinsic, $N>1$ observable, and $N>1$ intrinsic correlations then provides a way to identify how the additional $\theta$ dependence of the many-body state modifies this baseline and how the two contributions combine in the full spinor-anyon correlation. Motivated by these limiting cases, we use the more general two-branch form
\begin{align}
C_{\mathrm{avg}}(r)
\sim
A_+\frac{e^{ik_+r}}{r^{\alpha_+}}
+
A_-\frac{e^{ik_-r}}{r^{\alpha_-}}
\label{eq:numerical_fit_form}
\end{align}
for the numerical analysis below, without fixing $A_\pm$, $k_\pm$, or $\alpha_\pm$ to the approximate expressions above.
\begin{figure*}
\centering
\includegraphics{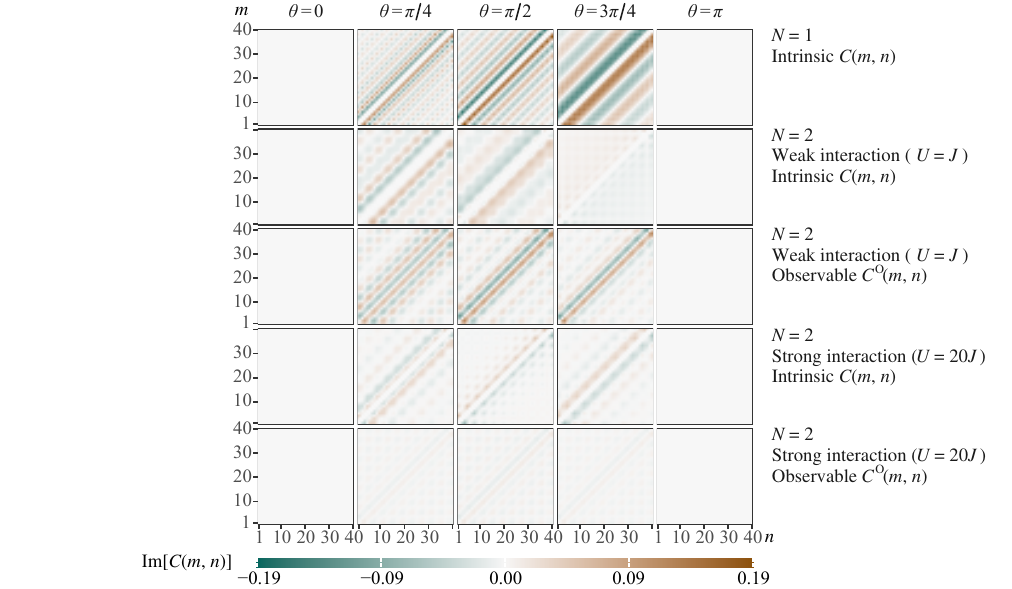}
\caption{\label{fig:obdmx}\textbf{Imaginary one-body density correlation reveals the real-space origin of momentum-space asymmetry.}
Imaginary part of the one-body density correlation for lattice filling $M/L=20/40$.
Columns correspond to statistical phases $\theta \in \{0,\;\pi/4,\;\pi/2,\;3\pi/4,\;\pi\}$.
From top to bottom, the rows show the intrinsic $N=1$ anyonic correlation, the intrinsic and observable $N=2$ correlations at weak interaction $U=J$, and the corresponding intrinsic and observable $N=2$ correlations at strong interaction $U=20J$.
The imaginary component vanishes at the fermionic and pseudobosonic limits and develops an antisymmetric off-diagonal structure at fractional exchange statistics.
Through the Fourier transform of the one-body density correlation, this imaginary component generates the $k\leftrightarrow -k$ asymmetry of the momentum distribution.
For $N=2$, the $\theta$ dependence of the intrinsic correlation $C(m,n)$ has two sources: the nonlocal statistical string in the anyonic operator and the $\theta$-dependent many-body state generated by occupation-dependent tunneling.
We separate these contributions by comparing three correlations.
The intrinsic $N=1$ correlation delineates the explicit string dependence, while the observable physical-fermion correlation $C^{\mathrm O}(m,n)$ for $N=2$, which contains no statistical string, delineates the state-induced contribution.
Comparison with the intrinsic $N=2$ correlation then reveals how these two mechanisms reinforce or compete to produce its antisymmetric imaginary structure and the associated momentum-space asymmetry.
Results are obtained using density-matrix renormalization group (DMRG) calculations with open boundary conditions.}
\end{figure*}
\begin{figure}
\centering
\includegraphics{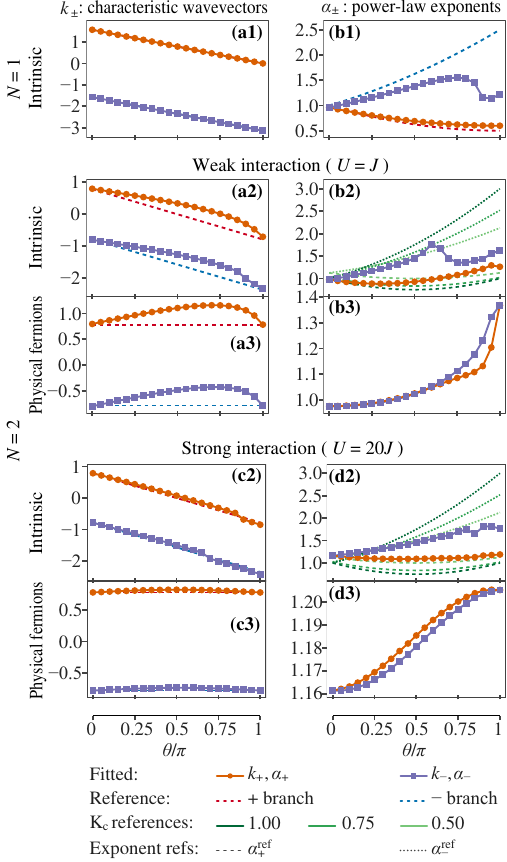}
\caption{\label{fig:N2fits}
\textbf{Characteristic wavevectors and power-law exponents separate the operator- and state-induced contributions to the one-body density correlations.}
Parameters extracted from the asymptotic analysis of the spatially averaged one-body density correlation for $L=40$, $M=20$, using DMRG with open boundary conditions.
The top row, as a reference, shows the intrinsic $N=1$ correlation; the lower rows show the intrinsic $N=2$ correlation and the observable physical-fermion $N=2$ correlation at the indicated interaction strength.
The left panel of each row shows the extracted characteristic wavevectors $k_+$ and $k_-$, while the right panel shows the corresponding decay exponents $\alpha_+$ and $\alpha_-$.
For $N=1$, the dashed reference curves are the spinless-anyon predictions
$k_\pm=(\pm1-\kappa)k_{\mathrm F}$ and
$\alpha_\pm=1\mp\kappa+\kappa^2/2$.
For the intrinsic $N=2$ correlation, dashed curves in the wavevector panels show the string-only reference
$k_+=(1-2\kappa)k_{\mathrm F}$ and
$k_-=-(1+2\kappa)k_{\mathrm F}$, while the exponent panels compare the numerical values with the approximate string-only forms for representative charge Luttinger parameters $K_{\mathrm c}=1$, $0.75$, and $0.5$.
For the observable $N=2$ correlation, the horizontal dashed lines at $\pm k_{\mathrm F}$ provide the fermionic reference wavevectors; its $\theta$ dependence originates solely from reconstruction of the mapped many-body state.
The comparison shows how the nonlocal statistical string and the $\theta$-dependent state modify the oscillation scales and algebraic decay separately, and how their combined effect determines the intrinsic $N=2$ correlation.
}
\end{figure}

\section{\label{sec:appendD}Origin of momentum-space asymmetry}
Here, we examine the origin of the momentum-distribution asymmetry in Fig.~\ref{fig:anyonsdis}, and connect it to the two sources of statistical-phase dependence. The comparison between $N=1$ and $N=2$ in this part is used only to separate these two contributions and should not be interpreted as an analysis of the dependence on the number of components. The $N$ dependence and its interplay with $\theta$ is considered through the numerical fits of the one-body density correlations in Appendix~\ref{sec:appendE}.

For this purpose, we examine the imaginary part of the one-body density correlation $C(m,n)$. As follows from Eq.~\eqref{eqn:brokensym}, $\Im[C(m,n)]$ generates the odd part of the momentum distribution and therefore directly signals inversion breaking. Figure~\ref{fig:obdmx} shows that $\Im[C(m,n)]$ vanishes at the symmetric limits $\theta=0$ and $\pi$, while fractional exchange statistics produces an antisymmetric, spatially oscillating structure.

To distinguish the two sources of this $\theta$ dependence, the $N=1$ intrinsic correlation is used to identify the contribution of the nonlocal statistical string, while the $N=2$ observable physical-fermion correlation separates the contribution associated with the $\theta$-dependent many-body state. Both exhibit oscillations, but with different characteristic periods and relative phases. The $N=2$ intrinsic correlation contains both effects. Their differing oscillation scales and phases allow the two contributions to compete and partially cancel, such that the imaginary part of the intrinsic correlation can be less pronounced than either contribution considered separately. At strong interaction, the observable contribution is strongly suppressed while the intrinsic correlation retains a finite imaginary component, indicating that the nonlocal string provides the dominant surviving source of inversion asymmetry. 

This real-space picture also clarifies the diagnostics introduced in Appendix~\ref{sec:appendB}. The mirror asymmetry $\mathcal A_{\mathrm M}$ directly reflects the finite odd component generated by $\Im[C(m,n)]$, while the mean momentum $\bar{k}$ measures the residual displacement after the positive- and negative-momentum contributions are combined. The partial cancellation between the string- and state-induced oscillations therefore provides a natural explanation for why the intrinsic distribution can retain a sizable $\mathcal A_{\mathrm M}$ while exhibiting a comparatively small $\bar{k}$.

\section{\label{sec:appendE} Interpreting \texorpdfstring{$N=2$}{N=2} momentum distributions}
Here, we use the one-body density correlations to connect the momentum-space diagnostics of Appendix~\ref{sec:appendB} with the real-space origin of the asymmetry identified in Appendix~\ref{sec:appendD}. The extracted characteristic wavevectors, decay exponents, and amplitudes quantify how the oscillation scales and long-distance weights of the correlation evolve with statistical phase and interaction strength.

For the parameters of Fig.~\ref{fig:obdmx}, a direct fit of Eq.~\eqref{eq:numerical_fit_form} would require the simultaneous determination of the six parameters $A_\pm$, $k_\pm$, and $\alpha_\pm$. For the finite open chains considered here, fitting all six parameters simultaneously is not sufficiently stable. Finite-size and boundary effects restrict the range over which the asymptotic form applies, and the correlation is only approximately described by the two dominant oscillatory branches. We thus extract characteristic wavevectors and decay exponents independently, presented in Fig.~\ref{fig:N2fits}, leaving only the amplitudes for the final fit. 

For $N=1$ at half filling, $k_{\mathrm F}=\pi/2$ and the two characteristic wavevectors satisfy $k_{-} - k_{+} = -\pi$. The characteristic wavevector is thus obtained from
\begin{align}
    k_+
    \simeq
    \frac{1}{2}\mathrm{Arg}\left[\sum_r
    C_{\mathrm{avg}}^*(r)
    C_{\mathrm{avg}}(r+2)\right],
    \quad
    k_-=k_+-\pi\,.
\end{align}
After removing the $k_+$ oscillation by $D(r) = e^{-ik_+ r}C_{\mathrm{avg}}(r)$, the two $\pm$ contributions are separated into slowly varying parts:
\begin{align}
    D_+(r)
    &=
    \frac{D(r-1)+2D(r)+D(r+1)}{4},
    \\
    D_-(r)
    &=
    \frac{(-1)^r[2D(r)-D(r-1)-D(r+1)]}{4}\,.
\end{align}
The decay exponents follow from
\begin{align}
    \ln\abs{D_\pm(r)}
    =
    \mathrm{const}
    -
    \alpha_\pm\ln r\,.
\end{align}
With $k_\pm$ and $\alpha_\pm$ fixed in this way, the remaining amplitudes $A_\pm$ are obtained from a linear fit of
\begin{align}
    C_{\mathrm{avg,fit}}(r)
    =
    A_+
    \frac{e^{ik_+r}}{r^{\alpha_+}}
    +
    A_-
    \frac{e^{ik_-r}}{r^{\alpha_-}}\,.
\end{align}

For $N=2$, we first obtain the characteristic wavevectors from a windowed Fourier transform of the complex correlation. The two characteristic features are identified from the extrema of the derivative of the Fourier spectrum. We then remove each oscillation in turn and apply a low-pass filter to obtain the corresponding slowly varying envelope. The decay exponents $\alpha_\pm$ are extracted from the log--log slopes of these envelopes. As for $N=1$, the extracted $k_\pm$ and $\alpha_\pm$ are then held fixed, and only the amplitudes $A_\pm$ are determined from the final reconstruction of $C_{\mathrm{avg}}(r)$. As $\theta$ approaches the pseudobosonic limit, the two characteristic features become less clearly separated, and the parameters extracted for the individual branches should therefore be interpreted with caution.

\textls[10]{The physics in the extracted parameters is that the two sources of $\theta$ act in opposite directions at weak interaction. For the $N=1$ intrinsic correlation, which isolates the nonlocal string, both characteristic wavevectors are displaced toward negative momentum as $\theta$ increases, while the two decay exponents separate: the $+$ contribution becomes longer ranged and the $-$ contribution more rapidly decaying. At larger $\theta$, however, the numerical value of $\alpha_-$ becomes less reliably determined, so its detailed $\theta$ dependence should be treated with caution, even though the ordering $\alpha_- > \alpha_+$ is retained.} The $N=2$ observable physical-fermion correlation shows the state-mediated effect alone and, for fractional exchange statistics ($0<\theta<\pi$), exhibits the opposite tendency at weak interaction: $k_+$ moves to larger momentum and $k_-$ becomes less negative, while both exponents increase moderately. Consequently, in the $N=2$ intrinsic correlation the state-mediated displacement partially opposes the explicit-string-dependent displacement; the extracted $k_{\pm}$ therefore deviate systematically from the string-only reference in the direction set by the observable correlation. This competition quantifies the real-space behavior identified in Appendix~\ref{sec:appendD}: the string- and state-induced contributions oscillate with different characteristic wavevectors and decay with different exponents, so that their imaginary parts can partially cancel in the intrinsic $N=2$ correlation. This accounts for the relatively small intrinsic mean momentum $\bar{k}$ despite a finite mirror asymmetry $\mathcal A_{\mathrm M}$. As for $N=1$, the extracted decay exponents $\alpha_\pm$ become less reliable at larger $\theta$ and should therefore be treated with caution.

The remaining diagnostics probe complementary aspects of the same reconstruction. Changes in $k_\pm$, $\alpha_\pm$, and $A_{\pm}$ alter both the even and odd parts of the momentum distribution and therefore contribute to the total redistribution $R$. In particular, changes in the decay exponents modify the sharpness of the associated momentum-space features, while changes in the relative amplitudes redistribute weight between them; together these effects are reflected in the observed change of momentum width $\Delta\sigma_k$. There is therefore no one-to-one correspondence between an individual fit parameter and a single diagnostic, but the two descriptions consistently separate displacement, asymmetry, redistribution, and broadening.

At strong interaction, the observable $N=2$ wavevectors remain close to $\pm k_{\mathrm F}$ and the corresponding exponents become only weakly dependent on $\theta$, consistent with the strong suppression of all four observable diagnostics in Appendix~\ref{sec:appendB}. The intrinsic correlation retains a much stronger $\theta$ dependence, particularly in the imbalance between its two contributions, consistent with the surviving intrinsic redistribution and mirror asymmetry. Close to the pseudobosonic limit, the characteristic Fourier features become poorly resolved and the parameters extracted for the weaker branch become less reliable and should be treated with caution.

\section{\label{sec:appendF}Symmetry constraints and \texorpdfstring{\\}{}short-time impurity dynamics}
Let us develop a more detailed understanding of the impact of the fractional exchange statistics on the evolution dynamics of an impurity. Consider three operators: inversion symmetry operator ($\mathcal{I}$) that maps site $j$ to $j'$ under the reflection about the system center, time-reversal operator $\mathcal{T}$ that gives the complex conjugation in the occupation basis, and the density-dependent gauge transformation $\mathcal{R} = \exp[-i \theta \sum_j n_j (n_j - 1) / 2]$. The Hamiltonian ($H_{\mathrm{F}}$) remains invariant under the action of the operator $\mathcal{K}=\mathcal{R}\mathcal{I}\mathcal{T}$, such that  $\mathcal{K}H_{\mathrm{F}}\mathcal{K}^{\dagger} = H_{\mathrm{F}}$. Using the combined symmetry $\mathcal{K}$, an inversion-symmetric initial state obeys $\expval{n_{j,\sigma}(t)}_{\theta} = \expval{n_{j',\sigma}(t)}_{-\theta}$ \cite{liu_anyon_2018}. Thus, inversion symmetry is enforced at $\theta=0$ and $\pi$ (and, in the noninteracting limit, at $U=0$), whereas for $0<\theta<\pi$ inversion-related propagation need not be equivalent. The sign of the resulting directional asymmetry is reversed under $\theta\rightarrow-\theta$.

To expose the microscopic origin of the asymmetric transport, we expand the time-evolution operator in powers of $t$ by considering a simple example. Let the initial state be inversion symmetric $\ket{\ldots,\; 0,\; \uparrow,\; \downarrow,\; \uparrow,\; 0,\; \ldots}$. Consider, after time evolution, two inversion-related final states: (\textit{a})~$\ket{\ldots,\; \uparrow,\; \downarrow,\; 0,\; \uparrow,\; 0,\; \ldots}$ and (\textit{b})~$\ket{\ldots,\; 0,\; \uparrow,\; 0,\; \downarrow,\; \uparrow,\; \ldots}$.

The perturbative expansion of the evolution operator is
\begin{align}
    \mathcal{U} = e^{-iH_{\mathrm{F}}t} = \sum\limits_{k=0}^{\infty} \dfrac{\left(-iH_{\mathrm{F}}t\right)^k}{k!}
\end{align}
The amplitude for the time evolution of the initial state to state (\textit{a}) can be obtained from the second-order terms as $J^2$ for the process $\ket{\ldots,\; 0,\; \uparrow,\; \downarrow,\; \uparrow,\; 0,\; \ldots} \rightarrow \ket{\ldots,\; \uparrow,\; 0,\; \downarrow,\; \uparrow,\; 0,\; \ldots} \rightarrow\; \ket{\ldots,\; \uparrow,\; \downarrow,\; 0,\; \uparrow,\; 0,\; \ldots}$ and as $J^2e^{-i\theta}$ for the process $\,\ket{\ldots,\; 0,\; \uparrow,\; \downarrow,\; \uparrow,\; 0,\; \ldots} \rightarrow \ket{\ldots,\; 0,\; \uparrow\downarrow,\; 0,\; \uparrow,\; 0,\; \ldots} \rightarrow \;\ket{\ldots,\; \uparrow,\; \downarrow,\; 0,\; \uparrow,\; 0,\; \ldots}$. In the presence of interaction, there is an additional contribution to the amplitude from the third-order term: $J^2Ue^{-i\theta}$.
 
\begin{figure}
\centering
\includegraphics{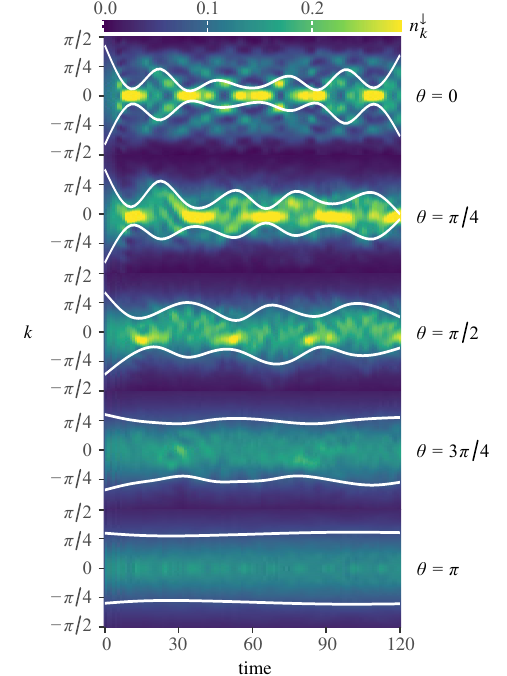}
\caption{\label{fig:iradd}\textbf{Fractional exchange statistics suppresses momentum-space breathing after impurity release.}
Impurity momentum distribution $n_k^\downarrow$ following release from the trap center at weak interaction $(U=J)$, corresponding to the real-space dynamics in Fig.~\ref{fig:irir}(a). Rows correspond to statistical phases $\theta \in \{0,\;\pi/4,\;\pi/2,\;3\pi/4,\;\pi\}$. The two white curves trace the generalized-additive-model-smoothed FWHM and together demarcate the breathing envelope. The breathing envelope oscillates progressively more slowly with increasing $\theta$ and becomes nearly stationary in the pseudobosonic limit, demonstrating the suppression of momentum-space breathing. Results are obtained for a weak harmonic trap $(\Omega=5\times10^{-3}J)$ using the time-dependent variational principle (TDVP) method for system size $L=21$, number of particles $M=10$, number of spin-up anyons $M^\uparrow=M-1$, and open boundary conditions. Colors saturate at the 99th percentile. Time is expressed in units of $J^{-1}$.}	
\end{figure}

Hence, the transition-amplitude magnitude for the evolution to state (\textit{a}) is 
\begin{align*}
S_a = \dfrac{t^2}{2}\abs{J^{2}\left(1+e^{-i\theta}\right)+\dfrac{-it}{3}J^{2}Ue^{-i\theta}}
\end{align*}
Similarly, for state (\textit{b}), it is
\begin{align*}
S_b = \dfrac{t^2}{2}\abs{J^{2}\left(1+e^{+i\theta}\right)+\dfrac{-it}{3}J^{2}Ue^{+i\theta}}
\end{align*}

The occupation-dependent tunneling assigns the conjugate statistical phases $e^{-i\theta}$ and $e^{+i\theta}$ to the two inversion-related processes. Their interference with the interaction-dependent contribution therefore produces different transition amplitudes, $S_a\neq S_b$, for $0<\theta<\pi$. Hence, anyonic transport is asymmetric. As $\theta$ increases from $0$, the transition amplitudes decrease, producing statistics-induced suppression of impurity transport. At $\theta=\pi$, the two second-order hopping paths interfere destructively and their leading contribution vanishes. The interaction-assisted third-order contribution remains finite but is identical for the two inversion-related processes, restoring inversion-symmetric propagation while strongly suppressing this short-time transport channel. This creates the dispersive to localized transport crossover and the asymmetry seen for fractional exchange statistics. The asymmetric transport is a direct outcome of the broken inversion symmetry in anyons.

\section{\label{sec:appendG}Impurity release: Momentum-space breathing} 
Here, we examine the momentum-space dynamics following the release of a trapped impurity from the center of the system. For fermions $(\theta=0)$, we had observed the momentum distribution periodically alternating between concentrating at small $\abs{k}$ and spreading toward larger $\abs{k}$, producing momentum-space breathing. To characterize this behavior directly, we determine the left and right half-maximum crossings of the dominant peak at each time step and interpolate between adjacent momentum points. The corresponding white curves in Fig.~\ref{fig:iradd} have a separation that is the full width at half maximum (FWHM) and it demarcates the resulting breathing envelope. The crossings are extracted from the full momentum distribution, while the heat map displays the central region $\abs{k}\leq\pi/2$, where the dynamics are of interest. Treating the two crossings separately also accommodates the asymmetric momentum distributions generated at fractional exchange statistics. The fermionic distribution exhibits pronounced oscillations of the envelope. These oscillation amplitudes become progressively weaker with increasing $\theta$ and are nearly absent at $\theta=\pi$, where the momentum distribution remains approximately stationary. Fractional exchange statistics therefore suppresses momentum-space breathing following impurity release. Statistics-dependent momentum-space breathing has also been obtained for strongly interacting two-component gases after a trap-frequency quench, including a many-body bounce that is absent from the density evolution \cite{patu_nonequilibrium_2023}. Here, the excitation is instead a local impurity release at finite interaction, and we find that the statistical phase progressively suppresses the observed breathing envelope.

\bibliography{references}
\end{document}